\documentclass[onecolumn, journal]{IEEEtran}

\usepackage{subfigure}
\usepackage{amsmath}
\usepackage{amsfonts}
\usepackage{mathrsfs}
\usepackage[dvips]{graphicx}

\usepackage{epsfig}
\usepackage{color}
\usepackage{graphicx}
\usepackage{array}
\usepackage{amssymb}
\usepackage{subfigure}

\newtheorem{thm}{Theorem}
\newtheorem{prop}{Proposition}
\newtheorem{lemma}{Lemma}
\begin{document}

\newcommand{\no}{\noindent}
\newcommand{\be}{\begin{eqnarray}}
\newcommand{\ee}{\end{eqnarray}}
\newcommand{\beeq}{\begin{equation}}
\newcommand{\eeeq}{\end{equation}}
\newcommand{\beeqs}{\begin{eqnarray*}}
\newcommand{\eeqs}{\end{eqnarray*}}
\newcommand{\bms}{\boldsymbol}
\newcommand{\expec}{\mathbf{E}}

\headheight 0in

\title{Jamming in Fixed-Rate Wireless Systems with Power Constraints - Part I: Fast Fading Channels}
\author{George T. Amariucai and Shuangqing Wei}
\maketitle
\footnotetext[1]{G. Amariucai and S. Wei are with the Department of ECE, Louisiana State
University. E-mail: gamari1@lsu.edu, swei@ece.lsu.edu.}

\begin{abstract}
This is the first part of a two-part paper that studies the problem of
jamming in a fixed-rate transmission system with fading, under the general
assumption that the jammer has no knowledge about either the codebook used
by the legitimate communication terminals, or the source's output. Both
transmitter and jammer are subject to power constraints which can be
enforced over each codeword (short-term / peak) or over all codewords
(long-term / average),  hence generating different scenarios. All our
jamming problems are formulated as zero-sum games, having the probability
of outage as pay-off function and power control functions as strategies.
The paper aims at providing a comprehensive coverage of these problems,
under fast and slow fading, peak and average power constraints, pure
and mixed strategies, with and without channel state information (CSI)
feedback.  In this first part we study the fast fading scenario. We
first assume full CSI to be available to all parties. For peak power
constraints, a Nash equilibrium of pure strategies is found. For average
power constraints, both pure and mixed strategies are investigated. With
pure strategies, we derive the optimal power control functions for
both intra-frame and inter-frame power allocation.  Maximin and
minimax solutions are found and shown to be different, which implies
the non-existence of a saddle point. In addition we provide alternative
perspectives in obtaining the optimal intra-frame power control functions
under the long-term power constraints. With mixed strategies, the Nash
equilibrium is found by solving the generalized form of an older problem
dating back to Bell and Cover \cite{bell}. Finally, for comparison
purposes, we derive a Nash equilibrium of the game in which no CSI is
fed back from the receiver. We show that full channel state information
brings only a very slight improvement in the system's performance.
\end{abstract}

\noindent {\bf \underline{Keywords:}}  Fast fading channels, fixed rate, $\lambda$-capacity, jamming, zero-sum game,
outage probability, power control.

\section{Introduction.}

The importance of designing anti-jamming strategies cannot be overstated, due to the
extremely wide deployment of wireless networks, the very essence of which makes them  vulnerable
to attacks. Although the bases of jamming and anti-jamming strategies have been set in the
80's and 90's \cite{basar3, basar1, medard}, new interest has been recently generated by the
increasing demand for wireless security. Jamming and anti-jamming strategies were developed for
the broadcast channel \cite{cioffi}, the multiple access channel \cite{uluk1}, and even
studied from the perspective of an arbitrarily varying channel \cite{hughes}. Under all
scenarios, the jamming problem is formulated as a two-player, zero-sum game. The corresponding
objective functions are the sum-rate \cite{cioffi}, the ergodic capacity \cite{uluk1} or the 
$\lambda$-capacity \cite{hughes}.
Although most often the jammer is assumed to have access to either the transmitter's
output or input \cite{basar3, medard, uluk2} and consequently is able to produce correlated
jamming signals,  the  correlation assumption can only be accurate for repeater protocols, or other
situations where the jammer gets the chance to jam a signal about which it has already obtained
some information from eavesdropping previous transmissions.

The approach of \cite{hughes} is quite relevant to our work. The jamming
problem is viewed as a special case of an arbitrarily varying channel
(AVC).  Constraints are placed either on the power invested in each
codeword (peak power constraints), or on the power averaged over all
codewords (average power constraints).  The $\lambda$-capacity, which is used
to evaluate system performance, is defined as the maximum transmission rate that
guarantees a probability of codeword error less than $\lambda$, under
random coding.  It is shown that when peak power constraints are imposed
on both transmitter and jammer, the $\lambda$-capacity is constant for
$0\leq\lambda<1$, and therefore is the same as the channel capacity.
No fading is assumed in \cite{hughes}, and consequently no power control
strategies are necessary.
However, fading channels are often the more practical models for wireless applications.

Traditionally, fast fading channels are characterized by their \emph{ergodic capacity}, which
is completely determined by the probability distribution of the channel coefficient and the
transmitter power constraints. The physical interpretation of this measure of channel quality
is related to the capabilities of channel codes. In the fast fading scenario, the codewords are
assumed long enough to reveal the long-term statistical properties of the fading coefficient (in
practical systems, this requirement may be satisfied by the use of interleaving \cite{tsevisw}).
Implicitly, power constraints are imposed over each codeword. Therefore, for achieving
asymptotic error free communication, all codewords need to be transmitted at the same rate
not exceeding the channel's ergodic capacity.

However, applications like video streams in multimedia often require fixed data rates
that could exceed the channel's ergodic capacity, but can tolerate non-zero codeword error
probabilities.  Therefore, in situations when the transmitter's available power is not
sufficient for supporting a certain rate for each codeword in the traditional framework,
the transmitter can choose to concentrate its power on transmitting only a subset of the
codewords, while dropping the others. This maneuver ensures error free decoding of the
transmitted messages, at the cost of a non-zero probability of message decoding error, which is
feasible when power constraints are imposed over the ensemble of all codewords, instead of over
each single codeword. This justifies the evaluation of fixed rate systems in fast fading
channels by a quantity that is best known to characterize slow fading channels: the \emph{outage
probability}. Note that unlike the case of slow fading, in fast fading channels, due to the large
codeword length, the channel conditions affecting the transmission of different codewords are
asymptotically identical.

\begin{figure}[]
\centering
\includegraphics[scale=0.5]{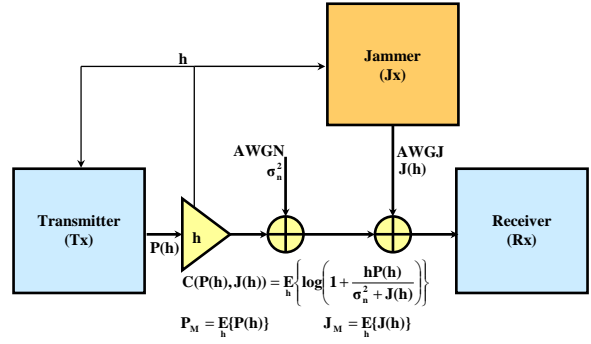}
\caption{Channel model}\label{channelmodel}
\end{figure}

In this paper, we consider a fast fading AWGN channel where codewords (we
denote the span of a codeword by the term \emph{frame}) are considered long enough to reveal
the long-term statistical properties of the fading coefficient. Our channel model is depicted  in
Figure \ref{channelmodel}. It was shown in \cite{caire2} that the ergodic capacity of the fast fading
AWGN channel can be achieved by a constant-rate, constant-power Gaussian codebook, provided that
when the fading coefficients are available at the transmitter, the transmitter employs a dynamic scaling
of the code symbols, by the appropriate power allocation function.
For this reason we assume in out model that the transmitter uses a capacity-achieving complex Gaussian codebook.
The jammer is assumed to have no knowledge about this codebook or the actual output of the transmitter,
and hence its most harmful strategy is to transmit white complex Gaussian noise \cite{diggavi}.

The channel coefficient is a complex number, the squared absolute value of which will be denoted
throughout this paper by $h$.
The average powers invested by the transmitter and jammer in transmitting and jamming a codeword, respectively,
are denoted by $P_M$ and $J_M$. The transmitter and the jammer are subject to either peak power constraints
(over each frame, or codeword) of the form $P_M\leq \mathcal{P}$ and $J_M\leq \mathcal{J}$ ,
or average power constraints (over all frames) of the form $\expec P_M\leq \mathcal{P}$ and $\expec J_M\leq \mathcal{J}$,
where the expectation is taken with respect to the players' strategies of allocating the powers $\mathcal{P}$
and $\mathcal{J}$ between frames.

A codeword is decoded with strictly positive probability of error (i.e. outage) if the ergodic capacity calculated over the
frame is below the fixed rate $R$. The probability of this event (the equivalent of $\lambda$ in \cite{hughes})
will be denoted as the \emph{probability of outage} $P_{out}$.
The transmitter aims at minimizing the probability of outage for a fixed rate $R$, while
the jammer attempts to maximize it. Our contributions can be summarized as below:

\begin{itemize}
\item  We first investigate the scenario where full channel state information (CSI) is available
to all parties. For this case we show that peak power constraints are not efficient for high rate transmissions or large
jammer power;

\item We formulate the scenario of average transmitter/jammer power
constraints as a two-person, zero-sum game with the probability of outage
as the pay-off function.

\item Under average power constraints, we first investigate pure strategies and find the maximin
and minimax solutions, as a result of two levels of power control: one within frames and one concerning the
additional randomization introduced by the transmitter. Optimal strategies
are derived for both levels, and it is shown that a Nash equilibrium of pure
strategies does not exist in general.

\item As a result, we investigate mixed strategies and find the (unique) Nash equilibrium by solving
a generalized version of a game that was first discussed by Bell and Cover \cite{bell} and then
extended by Hughes and Narayan \cite{hughes}.

\item Finally, for comparison purposes, we find the optimal transmitter and jammer mixed strategies for the
case when the receiver does not feed back the CSI. Our results show that  CSI feedback only brings slight
improvements in the overall transmission quality. 
\end{itemize}

One comment is in order. Note that Nash equilibria of mixed strategies are not always the best approach to practical jamming
situations. An equilibrium of mixed strategies usually assumes that none of the two players knows
exactly when or with what power the other player is going to transmit. While this may generally be
true for the legitimate transmitter, a smart jammer might constantly eavesdrop the channel and detect
both the legitimate transmitter's presence and its power level. Therefore, many real jamming
scenarios might be more accurately characterized by the solutions of the \emph{maximin problem formulation
with pure strategies} when the jammer tries to minimize and the transmitter tries to maximize the objective, and
the solutions of the \emph{minimax problem formulation with pure strategies} when the jammer tries to maximize
and the transmitter tries to minimize the objective (the latter case applies to the present paper).
At worst, these solutions provide a valid lower bound on system performance.

The paper is organized as follows. Section~\ref{section1} formalizes
the peak power constrained problem when full CSI is available to all parties.
It turns out that this problem has an intuitive solution.
Under the same full CSI assumption, Section \ref{section2} studies the problem of average power
constraints and pure strategies, and is divided into three subsections. The first one 
presents the optimal strategies for allocating power over one frame.
Using the results therein, the maximin and minimax solutions are derived
in Subsection \ref{ss4}. Some numerical results are shown in Subsection \ref{ss5}.
Section \ref{section3} investigates the problem of full CSI, average power
constraints and mixed strategies and provides the Nash equilibrium point.
The scenario when the channel coefficients are only known to the receiver is investigated in
Section \ref{section4}. Finally, conclusions are drawn in Section \ref{section5}.


\section{CSI Available to All Parties. Jamming Game with Peak Power Constraints.}\label{section1}

This game represents a more general version of the game discussed in Section IV.B of \cite{uluk1},
and its solution relies on the results therein.
The transmitter's goal is to:
\begin{gather}\label{game11}
\left\{ \begin{array}{cc} \textrm{Minimize} & \mbox{Pr}(C(P(h),J(h))<R)\\
\textrm{Subject to} & P_M=\expec_{h}[P(h)]\leq \mathcal{P},
\end{array} \right.
\end{gather}
while the jammer's goal is to:
\begin{gather}\label{game12}
\left\{ \begin{array}{cc} \textrm{Maximize} & \mbox{Pr}(C(P(h),J(h))<R)\\
\textrm{Subject to} & J_M=\expec_{h}[J(h)]\leq \mathcal{J}, 
\end{array} \right.
\end{gather}
where
\beeqs
C(P(h),J(h))=\expec_{h} \left[\log\left(1+\frac{hP(h)}{\sigma_N^2+J(h)} \right)\right].
\eeqs
is the ergodic capacity, which is completely determined by the p.d.f. of the channel coefficient $p(h)$ and the transmitter/jammer
power control strategies $P(h)$ and $J(h)$.
The expectation is defined as $\expec_h[f(h)]=\int_{h}f(h)p(h)dh$.

We prove that this game is closely related to the two player, zero-sum game of \cite{uluk1},
which has the mutual information between Tx and Rx as cost/reward function:

\begin{gather} \label{game21}
\textrm{Tx}\left\{ \begin{array}{cc} \textrm{Maximize} & C(P(h),J(h))\\
\textrm{Subject to} & P_M\leq \mathcal{P},
\end{array} \right.
\end{gather}

\begin{gather}\label{game22}
\textrm{Jx}\left\{ \begin{array}{cc} \textrm{Minimize} & C(P(h),J(h))\\
\textrm{Subject to} & J_M\leq \mathcal{J}. 
\end{array} \right.
\end{gather}

This latter game is characterized by the following proposition, proved in Section IV.B of \cite{uluk1}:
\vspace*{4pt}
\begin{prop}\label{prop_short_term}
The game of (\ref{game21}) and (\ref{game22}) has a Nash equilibrium point given by the following strategies:

\begin{gather}\label{sol11}
P^*(h)=\left\{ \begin{array}{ccc} \left[\frac{1}{\lambda}-\frac{\sigma_N^2}{h}\right]_+ & \textrm{if} & h<\frac{\sigma_N^2 \lambda}{1-\sigma_N^2 \nu}\\
\frac{h}{\lambda(h+\frac{\lambda}{\nu})} & \textrm{if} & h\geq\frac{\sigma_N^2 \lambda}{1-\sigma_N^2 \nu}
\end{array} \right.
\end{gather}

\begin{gather}\label{sol12}
J^*(h)=\left\{ \begin{array}{ccc} 0 & \textrm{if} & h<\frac{\sigma_N^2 \lambda}{1-\sigma_N^2 \nu}\\
\frac{h}{\nu(h+\frac{\lambda}{\nu})}-\sigma_n^2 & \textrm{if} & h\geq\frac{\sigma_N^2 \lambda}{1-\sigma_N^2 \nu}
\end{array} \right.
\end{gather}
 where $\lambda$ and $\nu$ are constants that can be determined from the power constraints and $[x]_+=\max\{x,0\}$.
\end{prop}
\vspace*{4pt}

The connection between the two games above is made clear in the following theorem, the proof of which
follows in the footsteps of \cite{caire} and is given in Appendix \ref{app1}.
\vspace*{4pt}
\begin{thm}\label{thm_short_term}
Let $P^*(h)$ and $J^*(h)$ denote the Nash equilibrium solutions of the game
described by (\ref{game21}) and (\ref{game22}).
Then the original game of (\ref{game11}), (\ref{game12}) has a Nash equilibrium point, which is given by
the following pair of strategies:

\begin{gather}
\widehat{P}(h)=\left\{ \begin{array}{ccc} P^*(h) & \textrm{if} & C(P^*(h),J^*(h))\geq R\\
P_a(h) & \textrm{if} &C(P^*(h),J^*(h))<R
\end{array} \right.
\end{gather}

\begin{gather}
\widehat{J}(h)=\left\{ \begin{array}{ccc} J_a(h) & \textrm{if} & C(P^*(h),J^*(h))> R\\
J^*(h) & \textrm{if} & C(P^*(h),J^*(h))\leq R,
\end{array} \right.
\end{gather}
where $P_a(h)$ and $J_a(h)$ are some arbitrary power allocations satisfying the respective power constraints.
(Note that no particular improvements are obtained by setting $P_a(h)=J_a(h)=0$, since only peak
power constraints are in effect.)
\end{thm}
\vspace*{4pt}

The results are intuitive: if the ergodic capacity under the optimal jammer/transmitter strategies
is larger than the fixed rate $R$, reliable
communication can be established over each frame, and hence the probability of outage is $P_{out}=0$.
In this case, the actual power allocation of the jammer does not matter anymore, since the jammer
has already lost the game.

On the other hand, if the ergodic capacity is less than $R$, outage occurs on all frames ($P_{out}=1$), and
the actual transmitter strategy makes no difference.
As will be shown in the next section, enforcing average power constraints in this case gives the transmitter
more freedom, and results in a smaller outage probability.


\section{CSI Available to All Parties. Jamming Game with Average Power Constraints: Pure Strategies.}\label{section2}

In this section power constraints are imposed over a large number of
frames rather than on each frame. The transmitter and jammer may increase
their transmission and jamming powers over any frame from $\mathcal{P}$
to $P_{M}$, and from $\mathcal{J}$ to $J_{M}$, respectively. To satisfy the
average power constraints imposed by $\mathcal{P}$ and $\mathcal{J}$, less power
has to be allocated to other frames. We shall prove that for both players, the optimal way to
control the power allocation between frames is to employ ON/OFF strategies.
Since all frames are equivalent from the point of view of their corresponding
channel realizations, the manner in which the ``discarded'' codewords are picked is
somewhat random.
However, note that this type of randomization
only aims at ensuring that a possibly larger $P_M$ or $J_M$ is obtained.
We don't consider mixing strategies in this section
\cite{meyerson}.  Although each player picks up a frame randomly, we
assume this is known by its opponent when considering the maxmin and
minimax problems as formulated below.
That is, the maximin scenario assumes the transmitter has perfect non-causal
access to the jammer's strategy (we say the jammer ``plays first''), while
the minimax case assumes the jammer has perfect, non-causal access to the
transmitter's strategy (we say the transmitter ``plays first'').
The first player in the minimax or maxmin cases is always more vulnerable
in the sense that the follower has the freedom to adapt its strategy such that
it minimizes the first player's payoff.

The minimax scenario is the more practical one. In addition to being pessimistic
from the system designer's point of view, it accurately models the situation where
the jammer (who is not interested in exchanging any information of its own) listens
to the feedback carrying the channel coefficients and senses the transmitter's presence
and power level, hence estimating the transmitter's strategy.
The maximin scenario is not of less importance, since it is required for determining the
non-existence of a Nash equilibrium and for comparison with the minimax approach.

An important remark should be made here. We shall prove in the sequel that under both the
pure strategies and the mixed strategies scenarios, the optimal power allocation over a frame is done
similarly. Therefore, the major difference between the two cases is in the strategies of allocating power to different frames.
We should note that it is easier for one of the players to detect the presence of the other player over
a frame, than to estimate the other player's transmission power. Under the minimax solution of pure
strategies, the jammer only needs to detect the presence of the transmitter (the optimal strategies are of
ON/OFF type) to have complete information about the transmitter's behavior. However, if the transmitter
chose to use mixed strategies, a complete characterization of its behavior would require not only
knowledge about its presence, but also about the power it decided to allocate to that frame.

The average power constrained jamming game can be formulated  as:
\begin{gather} \label{game31}
\textrm{Tx}\left\{ \begin{array}{cc} \textrm{Minimize} &
\mbox{Pr}(C(P(h),J(h))<R)\\
\textrm{Subject to} & E[P_M] \leq \mathcal{P}
\end{array} \right.
\end{gather}

\begin{gather} \label{game32}
\textrm{Jx}\left\{ \begin{array}{cc} \textrm{Maximize} &
\mbox{Pr}(C(P(h),J(h))<R)\\
\textrm{Subject to} & E[J_M]\leq \mathcal{J}
\end{array} \right.
\end{gather}
where $P_M$ and $J_M$ are defined as in (\ref{game11}), (\ref{game12}), the expectation is taken over all
frames with respect to the power allocation strategies introduced by the transmitter and jammer, and
$\mathcal{P}$ and $\mathcal{J}$ are the upper-bounds on average transmission
power of the source and jammer, respectively.


\subsection{Power Allocation within a Frame}\label{ss3}

The game between transmitter and jammer has two levels. The first
(coarser) level is about power allocation between frames, and has
the probability of outage as a cost/reward function.
The probability of outage is determined by the number of frames over which the transmitter is
not present or the jammer is successful in inducing outage.
This set is established in the first level of power control which is investigated in detail in
the next two subsections, but which cannot be derived
before the second level strategies are available.

The second (finer) level is that of power allocation  within a frame.
In this subsection we derive the optimal second level of power allocation
strategies for both maximin and minimax problems, and show they are connected
by a special kind of duality.

Note that decomposing the problem into several (two or three) levels and solving each one separately does not
restrict the generality of our solution. Our proofs are of a contradictory type. Instead of
directly deriving each optimal strategy, we assume an optimal solution has already been reached and
show it has to satisfy a set of properties. We first assume these properties are not satisfied,
and then show that under this assumption there is room for improvement. Thus we prove that any
solution not satisfying our set of properties cannot be optimal (i.e. the properties are necessary).
We pick the properties in such a manner that they are sufficient for the complete characterization of the
optimal solution. That is, we make sure that the system of necessary properties has a unique solution.

In the maximin case (when jammer plays first), assume that the
jammer has already allocated some power $J_M$ to a given frame.
Depending on the value of $J_M$, and its own
power constraints, the transmitter decides whether it wants to
achieve reliable communication over that frame. If it decides to
transmit, it needs to spend as little power as possible (the
transmitter will be able to use the saved power for achieving
reliable communication over another set of frames, and
thus to decrease the probability of outage). Therefore, the
transmitter's objective is to minimize the power $P_M$ spent for
achieving reliable communication over each frame.
Note that if the jammer is present over a frame, the value of $P_M$ required to achieve reliable
communication over that frame is a function of $J_M$.
However, the transmitter should attempt to minimize the required $P_M$ even when the jammer is absent.
The jammer's objective is then to allocate the given power $J_M$ over the frame
such that the required $P_M$ is maximized.

In the minimax scenario (when transmitter plays first)
the jammer's objective is to minimize the power $J_M$ used for jamming the transmission over a given
frame. The jammer will only transmit if the transmitter is present with some $P_M$.
The transmitter's objective is to distribute $P_M$ within a frame
such that the power required for jamming is maximized.

The two problems can be formulated as follows:

\vspace*{4pt}
{\bf \emph{Problem 1}} (for the maximin solution - jammer plays first)
\begin{gather}
\max_{J(h)\geq0} \Big[ \min_{P(h)\geq0} P_M=\expec_{h}[ P(h)], ~\textrm{s.t.}~  C(P(h),J(h))\geq R \Big]
\nonumber\\
\textrm{s.t.} ~\expec_{h}\left[J(h)\right]\leq J_M  ;
\label{probl1}
\end{gather}
\vspace*{4pt}

{\bf \emph{Problem 2}} (for the minimax solution - transmitter plays first)
\begin{gather}
\max_{P(h)\geq0} \Big[ \min_{J(h)\geq0} J_M=\expec_{h}[J(h)], ~\textrm{s.t.}~  C(P(h),J(h))\leq R \Big]
\nonumber\\
\textrm{s.t.} ~\expec_{h}[P(h)]\leq P_M.
\label{probl2}
\end{gather}
\vspace*{4pt}

Let $\mathfrak{m}$ denote the probability measure introduced by the probability density
function (p.d.f.) of $h$, i.e., for a set
$\mathscr{A}\subseteq \mathbb{R}_+$, we have
$\mathfrak{m}(\mathscr{A})=\int_{\mathscr{A}} p(h)dh$.
Denote $x(h)=J(h)+\sigma_N^2$.
Note that the expectation is defined as $\expec_h[f(h)]=\int_{h}f(h)p(h)dh$.
Similarly, we define $\expec_{h\in \mathscr{X}}[f(h)]=\int_{h\in \mathscr{X}}f(h)p(h)dh$.

\vspace*{4pt}
{\bf \emph{Solution of Problem 1}}\vspace*{4pt}

The transmitter's optimization problem:
\begin{gather}\label{pppxxpp}
\min_{P(h)\geq0} \expec_h [P(h)], ~\textrm{s. t.}~
\expec_h\left[\log \left( 1+\frac{h P(h)}{\sigma_N^2 +J(h)}
\right)\right] \geq R
\end{gather}
has linear cost function and convex constraints.
Write the Lagrangian as:
\begin{gather}
\mathbf{L_1}=\expec_h[P(h)]-\lambda \left\{\expec_h\left[\log \left( 1+\frac{h P(h)}{\sigma_N^2 +J(h)}
\right)\right] - R\right\}.
\end{gather}

With the notation $c= \exp (R)$ , the resulting KKT conditions yield the unique solution \cite{bertsek}:
\begin{gather}\label{Pn_first_expr}
P(h)=\left[\lambda -\frac{x(h)}{h}\right]_{+},~ h\in \mathbb{R}_+,
\end{gather}
where
\begin{gather}\label{exprlambda}
\lambda=c^{\frac{1}{\mathfrak{m}(\mathscr{M'})}}\left\{ \exp\left[\expec_{h\in \mathscr{M'}}\left(
\log \frac{x(h)}{h}\right)\right]\right\}^{\frac{1}{\mathfrak{m}(\mathscr{M'})}},
\end{gather}
and $\mathscr{M'}\subset \mathbb{R}_+$ is the set of channel coefficients
over which $\lambda\geq x(h)/h$, and $[z]_{+}=\max \{z,0\}$.
We say the transmitter is ``non-absent'' over $\mathscr{M'}$, and ``absent'' on $\mathbb{R}_+\setminus\mathscr{M'}$.

The following proposition, the proof of which is given in Appendix \ref{app6},
states that the jammer should only be present where the transmitter is non-absent.
\vspace*{4pt}
\begin{prop}\label{prop_lt1}
The jammer should only transmit where the transmitter is ''non-absent''.
Otherwise, if $J(h)>0$ and $\lambda<x(h)/h$ for $h$ in some set $\mathscr{S}\subset \mathbb{R}_+$,
the jammer can decrease $J(h)$ over $h \in \mathscr{S}$ and maintain the same required transmitter power over the frame.
\end{prop}
\vspace*{4pt}
Substituting (\ref{exprlambda}) in (\ref{pppxxpp}), the jammer's problem can be formulated as:
\begin{gather}
\textrm{Find}~ \max_{x(h)\geq \sigma_N^2}c^{\frac{1}{\mathfrak{m}(\mathscr{M'})}}
\mathfrak{m}(\mathscr{M'})\cdot\nonumber\\
\cdot\left\{ \exp \left[\expec_{h\in\mathscr{M'}}
\left(\log\frac{x(h)}{h}\right)\right]\right\}^{\frac{1}{\mathfrak{m}(\mathscr{M'})}}
-\expec_{h\in\mathscr{M'}}\left(\frac{x(h)}{h}\right)
\end{gather}
\begin{gather}\label{kkt1}
\textrm{subject to} ~\expec_h[x(h)]\leq (J_M+\sigma_N^2)
\end{gather}

Since the set $\mathscr{M'}$ depends on the jammer power allocation $J(h)$, solving
the optimization problem above analytically is difficult.
This is why we next provide an alternative method for finding the solution.
Our method examines the properties of the sets $\mathscr{M'}$ over
which the transmitter is present and $\mathscr{M''}$ over which the jammer is present,
as well as those of the optimal transmitter/jammer strategies.

Fixing $\mathscr{M'}$,
the Lagrangian for the jammer's optimization problem can be written as
\begin{gather}
\mathbf{L_2}=-P_M+\mu \left\{\expec_h[x(h)]-(J_M+\sigma_N^2)\right].
\end{gather}
This yields the new KKT conditions:
\begin{eqnarray} \label{kkt0}
\frac{1}{x(h)}\left\{\exp \left[\expec_{h\in\mathscr{M'}}\left(\log \frac{x(h)}{h}\right)\right]\right\}^
{\frac{1}{\mathfrak{m}(\mathscr{M'})}}c^{\frac{1}{\mathfrak{m}(\mathscr{M'})}}-{}\nonumber\\
{}-\frac{1}{h}-\mu =0
~\textrm{for}~h\in \mathscr{M''}, 
\end{eqnarray}
\begin{gather}\label{kkt11}
\expec_{h\in \mathscr{M''}}x(h)=J_M+\sigma_N^2\mathfrak{m}(\mathscr{M''}),
\end{gather}
\begin{gather}\label{kkt111}
\mu \geq 0,
\end{gather}
where $\mathscr{M''}$ is the set of channel coefficients on which the jammer transmits non-zero power.

For fixed $\mathscr{M'}$ and $\mathscr{M''}$, the jammer's optimal strategy has to satisfy these KKT conditions.
The resulting optimal strategy is
\begin{gather} \label{xn}
x(h)=\frac{h}{1+\mu h}\left\{c\exp \left[\expec_{h\in\mathscr{M'}}\left(\log \frac{x(h)}{h}\right)\right]\right\}^
{\frac{1}{\mathfrak{m}(\mathscr{M'})}}.
\end{gather}

The expression above states that for any two channel realizations with coefficients $h_i,~h_j$ belonging to $\mathscr{M''}$, we have
\begin{gather}
\frac{x(h_i)}{h_i}\geq\frac{x(h_j)}{h_j} \Leftrightarrow h_i\leq h_j \Leftrightarrow x(h_i)\leq x(h_j).
\end{gather}
Note that for any two channel realizations $h_i,~h_j \notin\mathscr{M''}$
(i.e. $x(h_i)=x(h_j)=\sigma_N^2$) we also have
\begin{gather}
\frac{x(h_i)}{h_i}\geq \frac{x(h_j)}{h_j} \Leftrightarrow h_i\leq h_j.
\end{gather}

\begin{figure}
\centering
\includegraphics[scale=1.0]{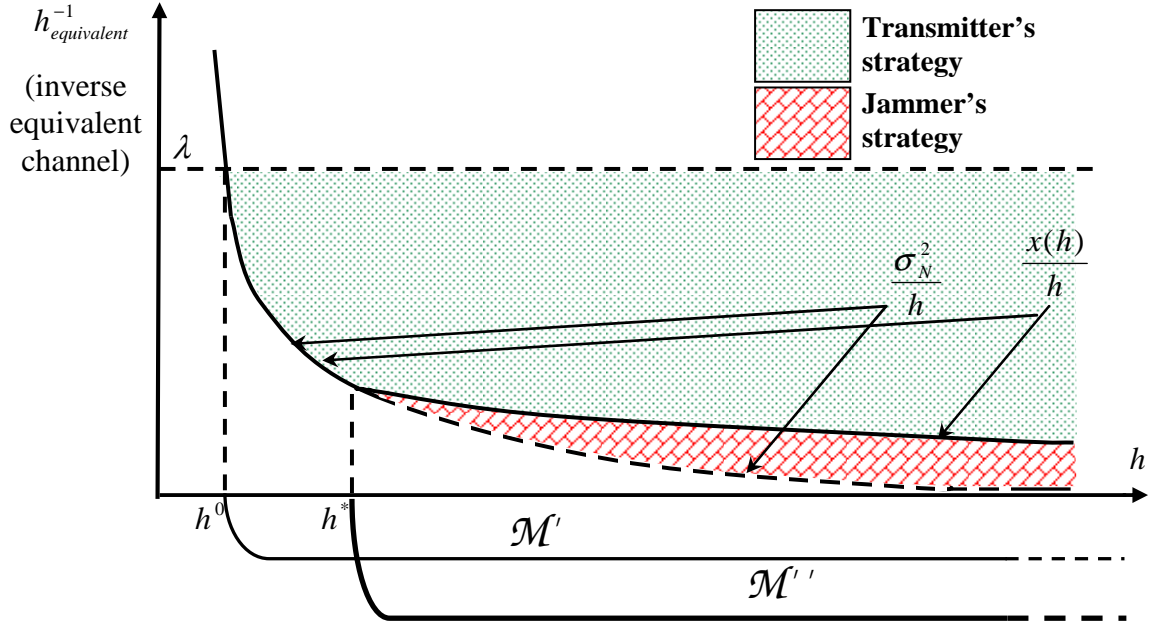}
\caption{Optimal second level power control strategies}\label{Minfp1}
\end{figure}

The following proposition brings more insight into the optimal jamming strategy.
Its proof is deferred to Appendix \ref{app7}.
\vspace*{4pt}
\begin{prop}\label{propmu}
The optimal jamming strategy is such that $x(h)/h$ is a continuous decreasing function of $h$
over all of $\mathbb{R}_+$, and $\mathscr{M''}$ is of the form $\mathscr{M''}=[h^*,\infty)$.
Moreover, this implies that $\mathscr{M'}$ is of the form $\mathscr{M'}=[h^0,\infty)$.
\end{prop}
\vspace*{4pt}
The optimal transmitter/jammer strategies for allocating power over a frame are described
in Figure \ref{Minfp1}.

Substituting (\ref{xn}) into (\ref{exprlambda}), we get a new expression for $\lambda$:
\begin{gather}\label{lambdanew}
\lambda=\frac{x(h)}{h}(1+\mu h), ~\textrm{for}~ h\in \mathscr{M''}
\end{gather}
which together with (\ref{Pn_first_expr}) yields
\begin{gather}\label{Pnlast}
P(h)=\mu x(h), ~\textrm{for}~ h\in \mathscr{M''}.
\end{gather}

An interesting remark which supports the results of the next subsection is that,
for the optimal solution of \emph{Problem 1}, $ \mu$ has to be strictly greater than zero,
hence eliminating the possibility that the jammer
allocates positive power to frames where the transmitter, although ``non-absent'', could allocate
zero power. In Appendix \ref{app7} it is shown how this remark follows from Proposition \ref{propmu}.

Taking expectation over $h\in \mathscr{M''}$ in (\ref{xn}), and using the constraint (\ref{kkt11}), we get
\begin{gather}\label{relxnx}
x(h)= \frac{J_M+\mathfrak{m}(\mathscr{M''})\sigma_N^2}{\frac{1+\mu h}{h}\expec_{h\in \mathscr{M''}}\frac{h}{1+\mu h}},
\end{gather}
for $h\in \mathscr{M''}$ and $x(h)=\sigma_N^2$ for $h\notin \mathscr{M''}$.

To solve for $\mu$, substitute (\ref{relxnx}) into (\ref{xn}):
\begin{eqnarray}\label{mu}
\left[\frac{J_M+\mathfrak{m}(\mathscr{M''})\sigma_N^2}{\expec_{h\in \mathscr{M''}}\frac{h}{1+\mu h}}\right]^
{\mathfrak{m}(\mathscr{M'})-\mathfrak{m}(\mathscr{M''})}={}\nonumber\\
{}=c\exp\left[ \expec_{h\in \mathscr{M''}}\left(\log\frac{1}{1+\mu h}\right)\right]\cdot\nonumber\\
\cdot\exp\left[\expec_{h\in \mathscr{M'}-\mathscr{M''}}\left(\log\frac{\sigma_N^2}{h}\right)   \right].
\end{eqnarray}

The second level power allocation solution for the maximin problem is thus completely determined by
the triple $(\mathscr{M'}, \mathscr{M''}, \mu)$, or equivalently by $(h^0, h^*, \mu)$.
By Proposition \ref{propmu} above, $x(h^*)=\sigma_N^2$ (by continuity in $h^*$), and $\lambda=\sigma_N^2/h^0$.
Rearranging these two relations, along with (\ref{mu}) in a more convenient form, we obtain the following system
of equations, which has to hold for any solution to our problem:

\begin{eqnarray}\label{sys1}
h^0=\frac{h^*}{1+\mu h^*},
\end{eqnarray}
\begin{eqnarray}\label{sys2}
\frac{J_M}{\sigma_N^2}=\int_{h^*}^{\infty}\left(\frac{\frac{h}{1+\mu h}}{\frac{h^*}{1+\mu h^*}}-1\right)p(h)dh,
\end{eqnarray}
\begin{eqnarray}\label{sys3}
R=\int_{\frac{h^*}{1+\mu h^*}}^{h^*}\log\left(h\frac{1+\mu h^*}{h^*}\right)p(h)dh-\nonumber\\
-\int_{h^*}^{\infty}\log\left(\frac{1}{1+\mu h}\right)p(h)dh.
\end{eqnarray}

The equations above lead to the following result:
\vspace*{4pt}
\begin{prop}\label{propuniq}
The solution of the maximin second level power allocation problem is unique.
\end{prop}
\vspace*{4pt}
\begin{proof}
It is easy to see that the right hand side of (\ref{sys2}) is a strictly decreasing function of $h^*$,
for fixed $\mu$, and a strictly decreasing function of $\mu$, for fixed $h^*$, while being equal to a constant.
Hence, for given $J_M$, (\ref{sys2}) yields $\mu$ as a strictly decreasing function of $h^*$.

Similarly, the right hand side of (\ref{sys3}) is a strictly decreasing function of $h^*$,
for fixed $\mu$, and a strictly increasing function of $\mu$, for fixed $h^*$, while being equal to a constant.
Hence, (\ref{sys3}) yields $\mu$ as a strictly increasing function of $h^*$.

Since (\ref{sys2}) and (\ref{sys3}) have to be satisfied simultaneously by any solution, the solution has to be unique.
\end{proof}
\vspace*{4pt}

Another insightful remark that follows from (\ref{sys1})--(\ref{sys3}) is that as $J_M$ increases, both
$\mu$ and $h^*$ should be decreasing.

The following proposition, characterizing the $P_M(J_M)$ function, is necessary for deriving the optimal 
power allocation between frames in the next section. The proof is deferred to Appendix \ref{app5}.
\vspace*{4pt}
\begin{prop}\label{propconcave}
Under the optimal maximin second level power control strategies,
the ``required'' transmitter power $P_M$ over a frame is a strictly increasing, unbounded and concave function of the power
$J_M$ that the jammer invests in that frame.
\end{prop}
\vspace*{4pt}

Throughout the remainder of this paper, we shall denote by $\mathscr{P}_M(J_M)$ the function that characterizes
the ``required'' transmitter power over a frame where the jammer invests power $J_M$, in the maximin case.

\vspace*{4pt}
{\bf \emph{Solution of Problem 2}} \vspace*{4pt}

To solve the minimax intra-frame power allocation problem by
using the same techniques as in \emph{Problem 1} turns out to be more difficult.
Instead we use the above solution of \emph{Problem 1} and show that
for both problems, the second level power allocation follows the same rules.
\vspace*{4pt}
\begin{thm}\label{circ_pr_thm}
If $J_{M,1}$ is the value used for the second constraint in
\emph{Problem 1} above, and $P_{M,1}$ is the resulting value of the cost/reward function,
then solving \emph{Problem 2} with $P_M=P_{M,1}$ yields the cost/reward function
$J_M=J_{M,1}$. Moreover, any pair of second level power allocation strategies that makes an optimal solution
of \emph{Problem 1}, should also make an optimal solution of \emph{Problem 2}, and this also holds conversely.
\end{thm}
\vspace*{4pt}
\begin{proof}
The result is a direct consequence of Theorem \ref{theoremgeneral} in Appendix \ref{app4},
if we denote $x=P(h)$, $y=J(h)$, $f(x)=\expec_h[P(h)]$, $g(y)=\expec_h[J(h)]$ and $h(x,y)=C(P(h),J(h))$.
\end{proof}
We shall denote by $\mathscr{J}_M(P_M)$ the function that characterizes the ``required'' jamming power
over a frame where the transmitter invests power $P_M$, in the minimax case.
By Theorem \ref{circ_pr_thm}, we have that $\mathscr{J}_M(\mathscr{P}_M(J_M))=J_M$ and $\mathscr{P}_M(\mathscr{J}_M(P_M))=P_M$.

\vspace*{4pt}
{\bf \emph{Further comments on the power control within frames}} \vspace*{4pt}

Although the second level optimal power allocation strategies for the maximin
and minimax problems coincide,
this result should not be associated to the notion of Nash equilibrium,
since the two problems solved above do not form a zero-sum game,
while for the game of (\ref{game31}) and (\ref{game32}), first level power control
strategies are yet to be investigated.

Instead, the result should be interpreted as a form of duality. In fact, a much stronger result can be
observed as a consequence of Theorem \ref{theoremgeneral}.
Namely, a similar ``duality'' property links \emph{Problem 1} and \emph{Problem 2} above
to the auxiliary problem of (\ref{game21}) and (\ref{game22}) appearing in the peak power constraints scenario.
This explains the resemblance between the solution of the peak power constraints auxiliary problem
(\ref{sol12}) and the solution of \emph{Problem1} (\ref{lambdanew}), (\ref{Pnlast}).

Also, this common solution implies that $P(h)=\mu (J(h)+\sigma_N^2)$ over the set $\mathscr{M''}$ of channel
realizations where both jammer and transmitter are present. Although the transmitter is also active
over the set of nonzero measure $\mathscr{M'}\setminus \mathscr{M''}$ as in Figure \ref{Minfp1}, under practical
conditions the measure $\mathfrak{m}(\mathscr{M'}\setminus \mathscr{M''})$ of this set is relatively small.
This is the reason why the $\mathscr{P}_M(J_M)$ curve appears to be linear (although it is not) in Figure \ref{PMJMcurve}
of the numerical results section.


\subsection{Power Allocation between Frames}\label{ss4}

{\bf \emph{The Maximin Solution}}
\vspace*{4pt}

In this subsection we present the first level optimal power allocation
strategies for the maximin problem.
Recall that all frames are equivalent in the sense that they are all characterized by
the same channel realizations (although not necessarily occurring in the same chronological order).

The maximin scenario assumes that the transmitter is completely aware of the jammer's power
control strategy (only pure strategies are considered in this section).
Given a jammer's strategy that allocates different jamming powers to
different frames, the optimal way of allocating the transmitter's power is always to ensure that
reliable communication is obtained on the frames that require the least amount of transmitter power. 
The jammer's optimal strategy (which is based solely on this knowledge about the transmitter's strategy)
is presented in the following theorem.

\vspace*{4pt}
\begin{thm}\label{thm3_long_term}
Under the maximin scenario it is optimal for the jammer to allocate the same amount of power
$J_M=\mathcal{J}$ to all frames.
\end{thm}
\vspace*{4pt}

\begin{proof}
The proof relies on the concavity of $\mathscr{P}_M(J_M)$.
Consider the optimal maximin inter-frame power allocation strategies.
Let $\mathscr{S}, \mathscr{X}$ denote the sets of frames over which the transmitter and the jammer
are present, respectively. Note that the jammer can itself compute the optimal transmitter strategy in response to
its own, and hence is fully informed of the transmitter's response.

We first look at the set of frames $\mathscr{S}$ where the transmitter is active.
Denote the power invested by the jammer in this set by $\mathcal{J}_{\mathscr{S}}$.
Note that $\mathcal{P}$ is the average ``required'' transmitter power over
$\mathscr{S}$. 

If the two players' strategies are both optimal, then by modifying the allocation of $\mathcal{J}_{\mathscr{S}}$ 
over the frames of $\mathscr{S}$, the new average ``required'' transmitter power over
$\mathscr{S}$ can only be less than or equal to $\mathcal{P}$.
In other words, if we denote by $j_M$ the generic power level allocated by the jammer to a frame in
$\mathscr{S}$, then
\be\label{maxprobmaximin1}
\mathcal{P}=\max_{j_M}\int_{\mathscr{S}} \mathscr{P}_M(j_M)dj_M
\ee
subject to
\be\label{maxprobmaximin2}
\int_{\mathscr{S}} j_M dj_M=\mathcal{J}_{\mathscr{S}}.
\ee

By writing the KKT conditions for the maximization problem in (\ref{maxprobmaximin1}) and (\ref{maxprobmaximin2}) above,
it is straightforward to see that, at an optimum, $\frac{d\mathscr{P}_M(j_M)}{dj_M}$ should be constant all over
$\mathscr{S}$.
Taking into account the fact that $\mathscr{P}_M(j_M)$ is concave, we have that a uniform jamming power allocation of
$\mathcal{J}_{\mathscr{S}}$ over $\mathscr{S}$ achieves this optimum. 

We next look at the set of frames $\mathscr{X} \setminus \mathscr{S}$ where the transmitter cannot afford to be active.
This means that the ``required'' transmitter power over $\mathscr{X} \setminus \mathscr{S}$ is greater than or equal to
$\mathscr{P}_M(\mathcal{J}_{\mathscr{S}})$, or equivalently, the power invested by the jammer is greater than or equal to
$\mathcal{J}_{\mathscr{S}}$. But since the jammer already knows the transmitter's strategy, investing more than
$\mathcal{J}_{\mathscr{S}}$ in any of the frames of $\mathscr{X} \setminus \mathscr{S}$ would be a waste.

Therefore, under the optimal  maximin inter-frame power allocation strategies, the jammer can invest the same amount of power
into all the frames of $\mathscr{X} \bigcup \mathscr{S}$ (which means $\mathscr{S}\subset \mathscr{X}$).

But since the transmitter decides to match the required transmitter power on $\mathscr{S}$, there can be no frames where
the jammer is not active, and hence $\mathscr{X}$ is the set of all frames.
\end{proof}
\vspace*{4pt}

The jamming power allocated to each frame is $J_M=\mathcal{J}$.
In this case the transmitter faces an indifferent choice space.
The power required for the transmitter to achieve reliable communication is $P_M(J_M)$.
Hence, the transmitter's optimal strategy is to randomly pick as many frames as possible and
allocate power $P_M(J_M)$ to each of them.
This is equivalent to saying the transmitter is present over a frame with probability $p_t$,
given by $p_t=\frac{\mathcal{P}}{P_M(\mathcal{J})}$.
The resulting probability of outage is now $P_{out}=1-p_t$.

Note that if $\mathcal{P}\geq P_M(\mathcal{J})$, the probability of outage
can be reduced to zero. This corresponds to the case when the ergodic capacity of the channel,
computed in the conventional way, with peak power constraints, is larger than the rate $R$.

\vspace*{4pt}
{\bf \emph{The Minimax Solution}}
\vspace*{4pt}

Theorem \ref{circ_pr_thm} showed that for the minimax problem the power allocation within a frame, as well as
the relationship between the total powers used by transmitter and receiver over a particular frame, are identical to
the maximin problem.
Hence, by rotating the $\mathscr{P}_M(J_M)$ plane, we get the characteristic $\mathscr{J}_M(P_M)$ curve for the minimax problem.

The minimax scenario assumes that the jammer knows exactly when and with what power level the transmitter transmits.
Given a transmitter's strategy that allocates different powers to
different (equivalent) frames, the optimal way of allocating the
jammer's power is such that outage is first induced on the frames
that require the least amount of jamming power. 

Under these conditions, the transmitter's optimal strategy is presented in the following theorem.
\vspace*{4pt}
\begin{thm}\label{thm4_long_term_M>1}
Under the minimax scenario it is optimal for the transmitter to transmit
over a maximum number of frames, with the same power $P_M$ that minimizes the
probability of outage.
\end{thm}
\vspace*{4pt}
\begin{proof}
The proof relies on the convexity of $\mathscr{J}_M(P_M)$.
Consider the optimal minimax inter-frame power allocation strategies, and
let $\mathscr{S}, \mathscr{X}$ denote the sets of frames over which the transmitter and the jammer
are present, respectively. It is clear in this scenario that $\mathscr{X}\subset\mathscr{S}$.

We first look at the set of frames $\mathscr{X}$ where the jammer is active.
Denote the power invested by the jammer in this set by $\mathcal{J}_{\mathscr{X}}$, and
the power invested by the transmitter by $\mathcal{P}_{\mathscr{X}}$.
Note that $\mathcal{J}_{\mathscr{X}}$ is the average ``required'' jamming power over
$\mathscr{X}$. 

If the two players' strategies are both optimal, then by modifying the allocation of $\mathcal{P}_{\mathscr{X}}$ 
over the frames of $\mathscr{X}$, the new average ``required'' jamming power over
$\mathscr{X}$ can only be less than or equal to $\mathcal{J}_{\mathscr{X}}$.
In other words, if we denote by $p_M$ the generic power level allocated by the transmitter to a frame in
$\mathscr{X}$, then
\be\label{maxprobmaximin3}
\mathcal{J}_{\mathscr{X}}=\max_{p_M}\int_{\mathscr{X}} \mathscr{J}_M(p_M)dp_M
\ee
subject to
\be\label{maxprobmaximin4}
\int_{\mathscr{X}} p_M dp_M=\mathcal{P}_{\mathscr{X}}.
\ee

From the KKT conditions for the maximization problem in (\ref{maxprobmaximin3}) and (\ref{maxprobmaximin4}) above,
we see that, at an optimum, $\frac{d\mathscr{J}_M(p_M)}{dp_M}$ should be constant all over
$\mathscr{X}$.
Taking into account the fact that $\mathscr{J}_M(p_M)$ is convex, we have that a uniform transmitter power allocation of
$\mathcal{P}_{\mathscr{X}}$ over $\mathscr{X}$ achieves this optimum. 

We should emphasize here that the above arguments hold \emph{under the assumption that the jammer is
active over the whole set $\mathscr{X}$}, i.e. when $\mathscr{J}_M(p_M)>0$ over $\mathscr{X}$.
Of course, the overall required jamming power is increased by increasing the transmitter power over some
frames of $\mathscr{X}$, while neglecting the others. But this action modifies the set $\mathscr{X}$ itself,
and thus the initial assumptions.

We next look at the set of frames $\mathscr{S} \setminus \mathscr{X}$ where the jammer cannot afford to be active.
This means that the ``required'' jamming power over $\mathscr{S} \setminus \mathscr{X}$ is greater than or equal to
$\mathscr{J}_M(\mathcal{P}_{\mathscr{X}})$, or equivalently, the power invested by the transmitter is greater than or equal to
$\mathcal{P}_{\mathscr{X}}$. But since the transmitter already knows the jammer's strategy, investing more than
$\mathcal{P}_{\mathscr{X}}$ in any of the frames of $\mathscr{S} \setminus \mathscr{X}$ would be a waste.

Therefore, under the optimal  maximin inter-frame power allocation strategies, the transmitter can invest the same amount of power
into all the frames of $\mathscr{S}$.
\end{proof}
\vspace*{4pt}

The frames over which the transmitter allocates the optimal $P_M$ can be chosen at random.
This is equivalent to the transmitter being active over
a frame with probability $p_t$ given by $p_t=\frac{\mathcal{P}}{P_M}$.
Searching for the optimal $P_M$ is equivalent to searching for the optimal $p_t$.

The jammer's strategy is to attack as many of the frames where the transmitter is present as possible.
In order to induce outage over these frames, the jammer needs to allocate $\mathscr{J}_M(P_M)$ to
each of them. This is equivalent to the jammer transmitting $\mathscr{J}_M(P_M)$ on a frame on which the transmitter is present,
with probability $p_j$ given by $p_j=\frac{\mathcal{J}}{p_t \mathscr{J}_M(P_M)}$.
Note that $p_j$ represents the conditional probability that the jammer transmits over a frame,
given that the transmitter is present over that frame.
Outage over a frame occurs in two circumstances: either the transmitter (and consequently also the jammer) decides to
ignore the frame, or the transmitter attempts to transmit the corresponding codeword, but the jammer is present
(and since this is the minimax scenario, it is also successful).

The resulting probability of outage is $P_{out}=(1-p_t)+p_j p_t$ or, only as a function of $P_M$:
\begin{gather}
P_{out}=(1-\frac{\mathcal{P}}{P_M})+\frac{\mathcal{J}}{\mathscr{J}_M(P_M)}.
\end{gather}
The transmitter finds the optimal value of $P_M$ as the argument that minimizes $P_{out}$ above.
A numerical approach should perform exhaustive search with the desired resolution in the interval
$[\mathcal{P},P_{M,max}]$, where $P_{M,max}$ can be set such that
$\forall P_M>P_{M,max}$ we have $P_{out}(P_{M})>1-\epsilon$ for a fixed $\epsilon$.
Since $P_{out}\to 1$ as $P_{M}\to \infty$ independently
of the $\mathscr{J}_{M}(P_{M})$ curve, such a finite bound $P_{M,max}$ exists for any $\epsilon$.

Note that if the $\mathscr{P}_M(J_M)$ curve is strictly concave, the jammer can never achieve an outage
probability $P_{out}=1$. This is because the transmitter can invest all its power over
a small enough set of frames, such that the jamming power required to jam all the frames in this set
exceeds the jammer's power budget. 
If however the probability measure $\mathfrak{m}$ is chosen such that
$\mathscr{P}_M(J_M)$ is an affine function of the form $P_M=P_{M,0}+1/\theta J_M$,
and furthermore if $\mathcal{J}\geq \theta(\mathcal{P}-P_{M,0})$,
then $\frac{\mathcal{J}}{\mathscr{J}_M(P_M)}\geq \frac{\mathcal{P}-P_{M,0}}{P_M-P_{M,0}}
\geq \frac{\mathcal{P}}{P_M}$ for all values of $P_M$, and
the probability of outage becomes $P_{out}=1$.


\subsection{Some Numerical Results}\label{ss5}

\begin{figure}[h]
\centering
\includegraphics[scale=0.5]{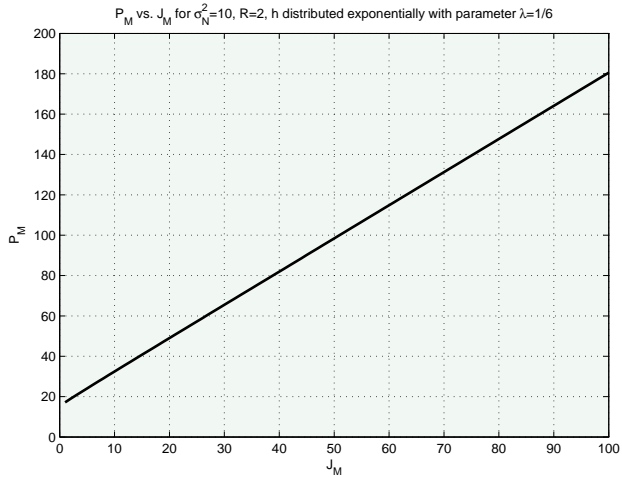}
\caption{$P_M$ vs. $J_M$ curve when $R=2$, $\sigma_N^2=10$ and $h$ is distributed exponentially, with
parameter $\lambda=1/6$.}\label{PMJMcurve}
\end{figure}

An example of the $\mathscr{P}_M(J_M)$ curve is given in Figure \ref{PMJMcurve} for a fixed rate $R=2$,
noise power $\sigma_N^2=10$ and a channel coefficient distributed exponentially, with parameter
$\lambda=1/6$.

\begin{figure}[h]
\centering
\includegraphics[scale=0.5]{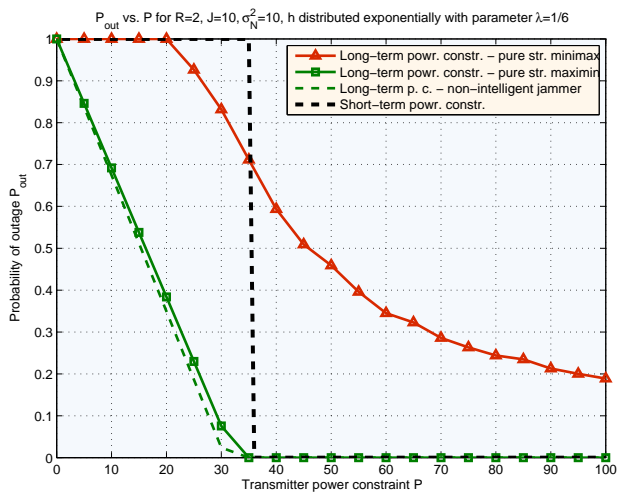}
\caption{Outage probability vs. transmitter power constraint $\mathcal{P}$ when $\mathcal{J}=10$,
 $R=2$, $\sigma_N^2=10$ and $h$ is distributed exponentially, with
parameter $\lambda=1/6$.}\label{PoutPfigure}
\end{figure}

For the same parameters used to generate Figure \ref{PMJMcurve}, the probability of outage was computed
for a jammer power constraint $\mathcal{J}=10$ and different values of the transmitter power constraint
$\mathcal{P}$. The results were plotted in Figure \ref{PoutPfigure}.
For comparison, the same figure shows $P_{out}(\mathcal{P})$ for the case when the jammer
does not use any power control strategy (non-intelligent jammer).
Since the jammer's first level of power control for the maximin scenario reduces to uniformly
distributing the available power to all frames, the only difference between the maximin
scenario and the non-intelligent jammer scenario is in the power allocation within frames.
However, as seen from Figure \ref{PoutPfigure}, this difference is almost negligible. 

\begin{figure}[h]
\centering
\includegraphics[scale=0.5]{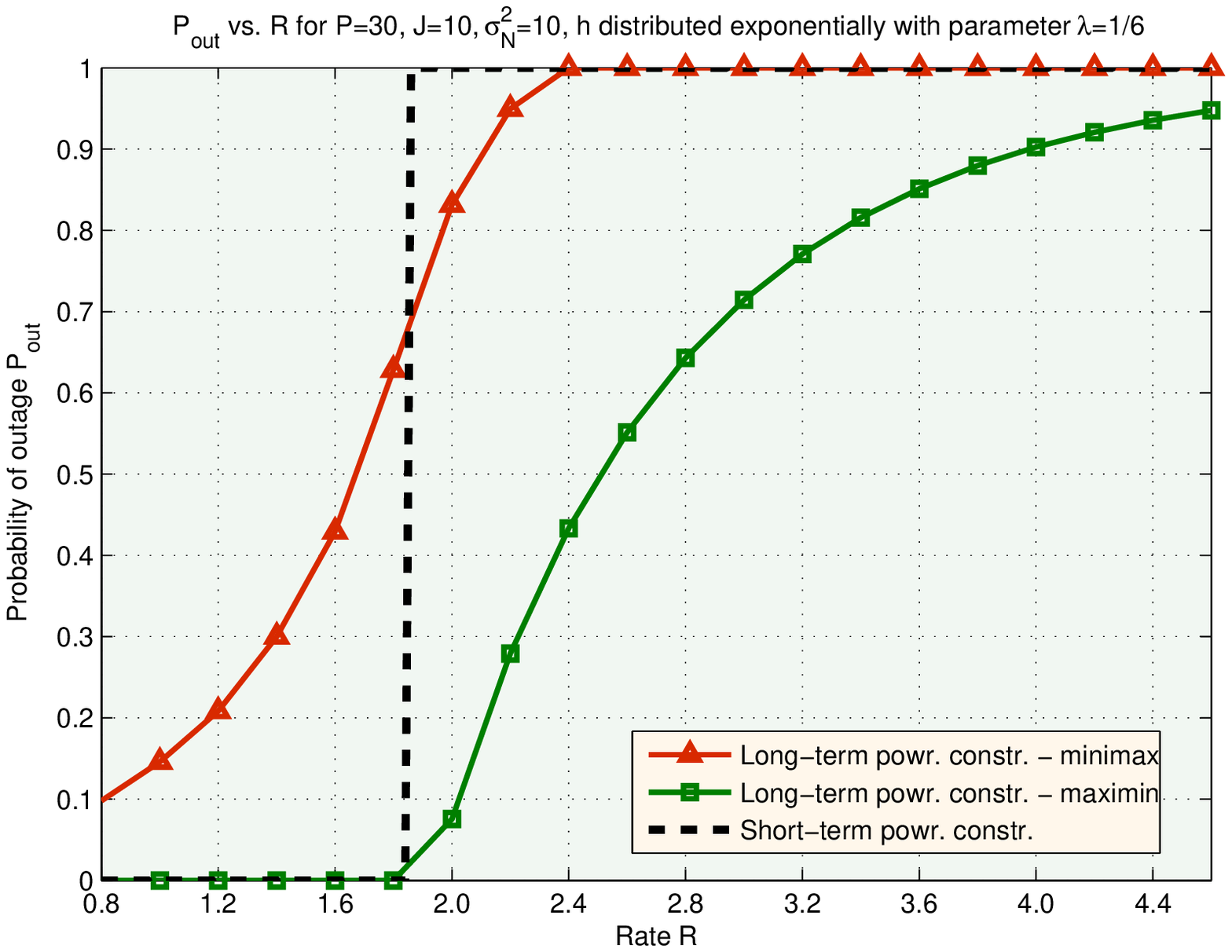}
\caption{Outage probability vs. rate for $\mathcal{P}=30$,
$\mathcal{J}=10$, $\sigma_N^2=10$ and $h$ is distributed exponentially, with
parameter $\lambda=1/6$.}\label{PoutRfigure}
\end{figure}

Figure \ref{PoutRfigure} shows how the outage probability varies with the rate $R$, for fixed
power constraints $\mathcal{P}=30$ and $\mathcal{J}=10$. The $P_{out}(R)$ curves delimitate
the achievable capacity vs. outage regions for both peak power constraints
and average power constraints (minimax and maximin cases).

Note that even for the minimax solution of the average power constraints problem, there exist
values of $\mathcal{P}$ (Figure \ref{PoutPfigure}), or of the rate $R$ (Figure \ref{PoutRfigure})
for which the outage probability is less than that achievable under peak power constraints.
  
Also note that the maximin curve coincides with the peak power constraints curve at large transmitter
power (in Figure \ref{PoutPfigure}) or at small rates (in Figure \ref{PoutRfigure}).
Recall that the jammer's strategy in the maximin scenario is the same as in the peak
power constraints scenario (i.e. the jammer allocates the same amount of power $\mathcal{J}$ to each frame).
Due to the favorable conditions in the regions characterized by large $\mathcal{P}$ or
small $R$, the transmitter can also spread its power uniformly over
all frames (just like in the peak power constraints scenario), overcoming the jammer completely
(hence the resulting zero probability of outage).


\section{CSI Available to All Parties. Jamming Game with Average Power Constraints: Mixed Strategies.}\label{section3}

In the previous section we studied the maximin and minimax solutions of the jamming game
when only pure strategies were allowed. Implicitly, we assumed that the power control strategies
employed by the first player are perfectly known to the second player, even if they include 
a form of ON/OFF randomization. We made a case that such a situation as the minimax case can emerge
when the jammer does not transmit unless it senses that the transmitter is on (and it can always
serve as a pessimistic scenario for the transmitter).

However,our previous assumption may sometimes be inappropriate from a practical point of view.
For example, if the transmitter does not stick with the optimal minimax solution, the jammer may have
a hard time following the transmitter's behavior. The reason for this is that, as we have already mentioned,
the jammer would find it much harder to correctly estimate the amount of power that the transmitter invests
in a given frame, than to just detect the presence of the transmitter.

In this section we investigate the jamming game with average power constraints when mixed (probabilistic)
strategies are considered. Similarly to the pure strategies scenario of the previous section, this game
is played on two levels, with the first (coarser) level dealing with power allocation between frames.
Its cost/reward function is the probability of outage. We assume that the jammer's and transmitter's
randomized strategies consist of picking the power values to be invested over a frame in a random
manner. In our previous notation, $P_M$ and $J_M$ are now random variables, and each frame is characterized by a
realization $(p_M,j_M)$ of the pair $(P_M,J_M)$.

Given this realization, each player has to distribute its power over the frame in an optimal way.
This is the purpose of the second (finer) level of power control. The objective of each player
at this level is to make the best of the available resources (i.e. the powers $(p_M,j_M)$).
This means maximizing (or minimizing, respectively) the average rate supported by the frame, in the hope that
the resulting average rate will be above (or below, respectively) the system's fixed rate $R$.


\subsection{Power allocation within a frame}\label{ss3_1}

We can formulate the second level of power control similarly to the two-player, zero-sum game of (\ref{game21})
and (\ref{game22}) having the ergodic capacity calculated over a frame $C(P(h),J(h))$ as cost function.
The difference is that under the current scenario, none of the players knows the other player's constraints,
because $(P_M,J_M)$ is a random event. Theorem \ref{thm1_long_term_mixed} below provides the optimal
transmitter/jammer strategies for power allocation within a frame.

\vspace*{4pt}
\begin{thm}\label{thm1_long_term_mixed}
Given a realization $(p_M,j_M)$ of $(P_M,J_M)$, let $\mathscr{P}_M(j_M)$ denote the solution of \emph{Problem 1}
in Section \ref{section2} with $J_M=j_M$, and $\mathscr{J}_M(p_M)$ denote the solution of \emph{Problem 2}
in Section \ref{section2} with $P_M=p_M$.

The transmitter's optimal strategy is the solution of the game
in (\ref{game21}) and (\ref{game22}), where the jammer is constrained to $\expec_h[J(h)]\leq \mathscr{J}_M(p_M)$ and
the transmitter is constrained to $\expec_h[P(h)]\leq p_M$.
The jammer's optimal strategy is the solution of the game in (\ref{game21}) and (\ref{game22}), where the
transmitter is constrained to $\expec_h[P(h)]\leq \mathscr{P}_M(j_M)$ and the jammer is constrained to
$\expec_h[J(h)]\leq j_M$.

Note that each of the two players deploys the strategy that results from the most pessimistic scenario that
it can handle successfully.
\end{thm}
\vspace*{4pt}
\begin{proof}
Denote the solution of the game in (\ref{game21}) and (\ref{game22}), where the jammer is constrained
to $\expec_h[J(h)]\leq \mathscr{J}_M(p_M)$ and the transmitter is constrained to $\expec_h[P(h)]\leq p_M$
by $(P_1(h),J_1(h))$, and the solution of the game in (\ref{game21}) and (\ref{game22}), where the
transmitter is constrained to $\expec_h[P(h)]\leq \mathscr{P}_M(j_M)$ and the jammer is constrained to
$\expec_h[J(h)]\leq j_M$ by $(P_2(h),J_2(h))$.

Denote the solution of the game in (\ref{game21}) and (\ref{game22}), where the jammer is constrained
to $\expec_h[J(h)]\leq j_M$ and the transmitter is constrained to $\expec_h[P(h)]\leq p_M$
by $(P_0(h),J_0(h))$..

By the duality property of Theorem \ref{theoremgeneral} in Appendix \ref{app4}, we must have
$C(P_1(h),J_1(h))=R$ and $C(P_2(h),J_2(h))=R$.

We will show that (i) even if mixed strategies are considered for the game in (\ref{game21}) and (\ref{game22}),
any Nash equilibrium has the same value as the Nash equilibrium of pure strategies; (ii) even if the jammer's power
$j_M$ is different from $\mathscr{J}_M(p_M)$, the transmitter's strategy is still optimal; (iii) even if the transmitter's power
$p_M$ is different from $\mathscr{P}_M(j_M)$, the jammer's strategy is still optimal. 

(i): Since the game of (\ref{game21}) and (\ref{game22}) is a two-person zero-sum game, all Nash equilibria of
mixed strategies yield the same value of the cost/reward function \cite{meyerson}.
Moreover, the two players are indifferent between all equilibria. It was shown in \cite{uluk1} that this game
has a Nash equilibrium of pure strategies. But any equilibrium of pure strategies is also an equilibrium of mixed strategies
\cite{meyerson} and hence it is enough to consider the equilibrium of pure strategies found in \cite{uluk1}.

(ii),(iii): Assume the transmitter plays the strategy given by $P_1(h)$. 

If $j_M=\mathscr{J}_M(p_M)$, it is clear that the optimal solution for both transmitter and jammer is the solution
of the game in (\ref{game21}) and (\ref{game22}), where the jammer is constrained
to $\expec_h[J(h)]\leq j_M$ and the transmitter is constrained to $\expec_h[P(h)]\leq p_M$. In this case,
it is as if each player knows the other player's power constraint.

If $j_M<\mathscr{J}_M(p_M)$, then by Lemma \ref{propapp51} in Appendix \ref{app5} we have that $J_0(h)<J_1(h)$. Since $C(P(h),J(h))$
is a strictly decreasing function of $J(h)$ (under the order relation defined in Appendix \ref{app4}),
this implies that $C(P_1(h),J_0(h))>R$. Note that $J_0(h)$ is the jammer's strategy when the jammer knows
the transmitter's power constraint $p_M$.
Thus we have shown that when the transmitter plays $P_1(h)$ and $j_M<\mathscr{J}_M(p_M)$, the jammer cannot induce outage
over the frame even if it knew the value of $p_M$.

The condition $j_M>\mathscr{J}_M(p_M)$ is equivalent to $p_M<\mathscr{P}_M(j_M)$ (by Theorem \ref{theoremgeneral}).
In this case, since the jammer plays the strategy given by $J_2(h)$, a similar argument as above
(but this time applied to the transmitter's strategy) shows that the transmitter cannot achieve
reliable communication over the frame even if it knew the exact value of $j_M$.

This accomplishes the proof and shows that $(P_1(h), J_2(h))$ is a Bayes equilibrium \cite{meyerson} for the game
with incomplete information describing the power allocation within a frame.
\end{proof}


\subsection{Power allocation between frames}\label{ss3_2}

Due to the form of the optimal second level power allocation strategies described in the previous subsection,
the outage probability can be expressed as
\be\label{alternpoutexpr}
P_{out}=Pr\{J_M\geq \mathscr{J}_M(P_M)\}={}\nonumber\\
{}=1-Pr\{P_M\geq \mathscr{P}_M(J_M)\},
\ee
where $\mathscr{P}_M(J_M)$ is the strictly increasing, unbounded and concave function of Proposition \ref{propconcave}.
The optimal mixed strategies for power allocation between frames are presented in the following theorem.

\vspace*{4pt}
\begin{thm}\label{thm2_long_term_mixed}
The unique Nash equilibrium of mixed strategies of the two-player, zero-sum game
with average power constraints described in (\ref{game31}) and (\ref{game32})
is attained by the pair of strategies $\left(F_P(p_M),F_J(j_M)\right)$ satisfying:
\be\label{gp1_011}
F_P(\mathscr{P}_M(y))\sim k_p\mathbb{U}([0,2v])(y)+(1-k_p)\Delta_0(y),
\ee 
\be\label{gp1_021}
F_J(\mathscr{J}_M(x))\sim k_j\mathbb{U}([0,J_M(2v)])(x)+(1-k_j)\Delta_0(x),
\ee 
where $\mathbb{U}([r,t])(\cdot)$ denotes the CDF of a uniform distribution
over the interval $[r,t]$, and $\Delta_0(\cdot)$ denotes the CDF of a Dirac distribution (i.e. a step function),
and the parameters $k_p,k_j\in [0,1]$ and $v\in [\max \{\mathcal{J}, \mathscr{J}_M(\mathcal{P})/2 \}, \infty)$ are uniquely
determined from the following steps:
\begin{enumerate}
\item Find the unique value $v_0$ which satisfies:
\be
\mathcal{P}\mathcal{J}=[\mathscr{P}_M(2v_0)-\mathcal{P}](2v_0-\mathcal{J}).
\ee
\item Compute $S(v_0)=\int_{0}^{2v_0}\mathscr{P}_M(y)dy-2v_0\mathcal{P}$.
\item If $S(v_0)<0$, then $v$ is the unique solution of 
\be
\int_{0}^{2v}\mathscr{P}_M(y)dy-2v\mathcal{P}=0,
\ee
\be
k_p=1
\ee
and 
\be
k_j=\frac{\mathcal{J}\mathscr{P}_M(2v)}{2v[\mathscr{P}_M(2v)-\mathcal{P}]}.
\ee
\item If $S(v_0)=0$ then $v=v_0$, $k_p=k_j=1$.
\item If $S(v_0)>0$, then $v$ is the unique solution of 
\be
\int_{0}^{2v}\mathscr{P}_M(y)dy-\mathscr{P}_M(2v)(2v-\mathcal{J})=0,
\ee
\be
k_p=\frac{2v\mathcal{P}}{\mathscr{P}_M(2v)[2v-\mathcal{J}]}
\ee
and 
\be
k_j=1.
\ee
\end{enumerate}
\end{thm}
\vspace*{4pt}
\begin{proof}
The proof follows directly from Theorem \ref{thm_gp1_1}  in Appendix \ref{app8},
by substituting $x=P_M$, $y=J_M$, $g(y)=\mathscr{P}_M(y)$, $g^{-1}(x)=\mathscr{J}_M(x)$,
$a=\mathcal{P}$ and $b=\mathcal{J}$. It is also interesting to note that the condition
$\int_0^{b}g(y)dy<\int_{g(b)}^{\infty}g^{-1}(x)dx$ is satisfied because $\mathscr{P}_M(y)$
is unbounded (Proposition \ref{propconcave}).
\end{proof}
\vspace*{4pt}


\subsection{Numerical results}\label{ss3_3}

\begin{figure}[h]
\centering
\includegraphics[scale=0.5]{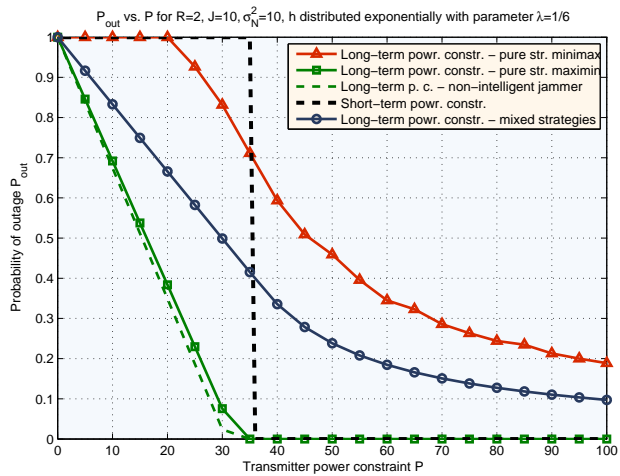}
\caption{Outage probability vs. transmitter power constraint $\mathcal{P}$ when $\mathcal{J}=10$,
 $R=2$, $\sigma_N^2=10$ and $h$ is distributed exponentially, with
parameter $\lambda=1/6$.}\label{fig_numericalresults3}
\end{figure}

For the same parameters as in subsection \ref{ss5} we evaluated numerically the optimal probabilistic
power control strategies.
Figure \ref{fig_numericalresults3} shows the probability of outage obtained under the mixed strategies
Nash equilibrium, versus the transmitter power constraint $\mathcal{P}$, for a fixed
rate $R=2$, noise power $\sigma_N^2=10$, a jammer power constraint $\mathcal{J}=10$
and a channel coefficient distributed exponentially, with parameter $\lambda=1/6$.
All the previously obtained curves are shown for comparison.
 
\begin{figure}[h]
\centering
\includegraphics[scale=0.5]{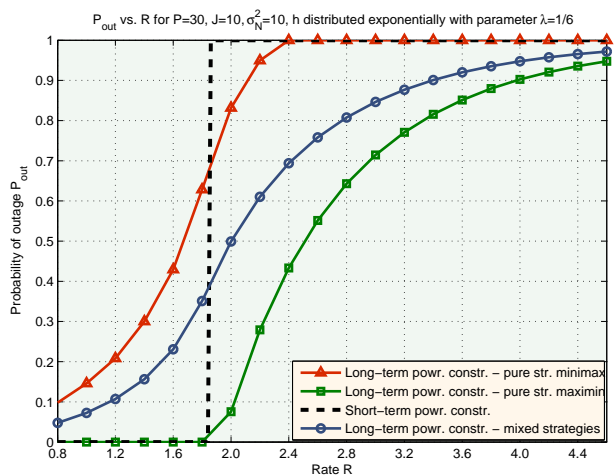}
\caption{Outage probability vs. rate for $\mathcal{P}=30$,
$\mathcal{J}=10$, $\sigma_N^2=10$ and $h$ is distributed exponentially, with
parameter $\lambda=1/6$.}\label{fig_numericalresults4}
\end{figure}

Figure \ref{fig_numericalresults4} shows the same probability of outage when $\mathcal{P}=30$ and the system
rate $R$ is varied.

In both figures it can be seen that the system performance under the Nash equilibrium of mixed strategies
is better (from the transmitter's point of view) than the minimax and worse than the maximin solutions
of the pure strategies game. This is expected since the pure strategies
solutions assume that the second player (the ``follower'') is constantly at a disadvantage with the first player
(the ``leader'').


\section{CSI Available to Receiver Only. Jamming Game with Average Power Constraints: Mixed Strategies}\label{section4}

In this section we investigate the scenario when the receiver does not feed back any channel state
information. Since we have already shown that the long term power constraints problem is the more
interesting and challenging one, we further focus only on the scenario of average power constraints and mixed strategies.
As in the previous sections, we have to discuss two levels of power control: within a frame
and between frames.

\subsection{Power allocation within a frame}

The jammer and transmitter powers allocated to each frame will be established in the next subsection. For now
we are concerned with the optimal power allocation within a frame, given the amounts of power invested in that frame
by each one of the players. For a given frame, denote these powers by $P_M$ and $J_M$, to be consistent with our
previous notation. Both the transmitter and the jammer will choose a probability distribution for the
randomly variable power levels $P$ and $J$, respectively, such that $\expec_{P} P\leq P_M$ and $\expec_{J} J \leq J_M$, 
where the notations $\expec_{P}$ and $\expec_{J}$ denote the expectations with respect to these probability distributions.
For the generic channel use, the channel coefficient $h$, the transmitter's power $P$ and the jammer's power $J$ are all
independent random variables, which yield the randomly variable instantaneous mutual information
$\log\left( 1+\frac{hP}{J+\sigma_N^2}\right)$. For a frame, this results in the ergodic capacity
$\expec_{h,P,J} \log\left( 1+\frac{hP}{J+\sigma_N^2}\right)$, where $\expec_{h}$ denotes expectation with
respect to the channel coefficient.

The transmitter's purpose is to use the allocated power $P_M$ in an attempt to make this ergodic capacity larger
than the rate $R$. Similarly, the jammer is concerned with using $J_M$ for making the ergodic capacity fall below $R$.
The problem of allocating the power within the frame can be written as:
\be
\max_{P:\expec_{P} P\leq P_M} \min_{J:\expec_{J} J \leq J_M} \expec_{h,P,J}
\log\left( 1+\frac{hP}{J+\sigma_N^2}\right).
\ee

Denote $L(P,J)=\expec_h \log\left( 1+\frac{hP}{J+\sigma_N^2}\right)$ and let us observe that
\be
\frac{dL}{dP}=\expec_h \frac{h}{Ph+J+\sigma_N^2}>0,
\ee
\be
\frac{dL}{dJ}=-\expec_h \frac{Ph}{(Ph+J+\sigma_N^2)(J+\sigma_N^2)}<0,
\ee
\be
\frac{d^2L}{dP^2}=-\expec_h \left(\frac{h}{Ph+J+\sigma_N^2}\right)^2<0,
\ee
\be
\frac{d^2L}{dJ^2}={}\nonumber\\
{}=\expec_h \frac{Ph(Ph+2J+2\sigma_N^2)}{[J^2+J(Ph+2\sigma_N^2)+\sigma_N^2(Ph+\sigma_N^2)]^2}>0,
\ee
which implies that $L(P,J)$ is a strictly increasing, concave function of $P$ for fixed $J$, and
a strictly decreasing, convex function of $J$ for fixed $P$.

Thus, we can write
\be
\expec_{h,P} \log\left( 1+\frac{hP}{J_M+\sigma_N^2}\right)\leq{}\nonumber\\
{}\leq \expec_{h} \log\left( 1+\frac{hP_M}{J_M+\sigma_N^2}\right)\leq{}\nonumber\\
{}\leq \expec_{h,J} \log\left( 1+\frac{hP_M}{J+\sigma_N^2}\right),
\ee
and hence the uniform distribution of $P_M$ and $J_M$ over the frame achieves a Nash equilibrium.
A frame to which the transmitter allocates power $P_M$ and the jammer allocates power $J_M$ is in outage
if and only if
\be\label{jhfscuyobeuircrb}
\expec_{h} \log\left( 1+\frac{hP_M}{J_M+\sigma_N^2}\right)\leq R.
\ee
The probability of this event depends on the power allocation between frames and is the subject of
the first level of power control treated in the next subsection.

But before we get to that, we need to make several comments. Note that if we force equality in (\ref{jhfscuyobeuircrb}) above,
we obtain a $\mathscr{P'}_M(J_M)$ curve as in Section \ref{section2}. 
It is straightforward to see that the $\mathscr{P'}_M(J_M)$ curve is affine, because solving
(\ref{jhfscuyobeuircrb}) with equality yields $P_M=\mu' (J_M+\sigma_N^2)$ where $\mu'$ is the (unique) solution of
$\expec_{h} \log\left( 1+\mu' h\right)= R$.
Recall that the curve $\mathscr{P}_M(J_M)$ of Section \ref{section2} (with full CSI) is \emph{almost} affine due
to the fact that the measure of the set of channel realizations, within a frame, over which the transmitter is
present but the jammer is not, is often quite small. For this reason, we expect the $\mathscr{P'}_M(J_M)$ and
the $\mathscr{P}_M(J_M)$ curves to be very close to each other.

Although the two curves are still different in general, they have the same physical interpretation: if the jammer
invests power $j_M$ over a frame, and the power $p_M$ invested by the transmitter satisfies $p_M<\mathscr{P'}_M(j_M)$,
then the frame is in outage.
Otherwise, if $p_M>\mathscr{P'}_M(j_M)$, the frame supports the asymptotically error-free decoding of the transmitted codeword.

As in Section \ref{section2}, we shall denote by $\mathscr{J'}_M(P_M)$ the ``inverse'' of the $\mathscr{P'}_M(J_M)$ function,
or the symmetric of the $\mathscr{P'}_M(J_M)$ curve with respect to the first bisector.


\subsection{Power allocation between frames}

The arguments of this subsection are very similar to those of Subsection \ref{ss3_2} and will not be discussed
in great detail. We have seen that the outage probability can be expressed as
\be\label{alternpoutexpr_nocsi}
P_{out}=Pr\{J_M\geq \mathscr{J'}_M(P_M)\}={}\nonumber\\
{}=1-Pr\{P_M\geq \mathscr{P'}_M(J_M)\},
\ee
where $\mathscr{P'}_M(J_M)$ is an affine, and hence strictly increasing and unbounded function
of the form $\mathscr{P'}_M(J_M)=\mu' J_M+\mu'\sigma_N^2$.
The optimal mixed strategies for power allocation between frames are presented in the following theorem.

\vspace*{4pt}
\begin{thm}\label{thm2_long_term_mixed_nocsi}
The unique Nash equilibrium of mixed strategies of our two-player, zero-sum game
with average power constraints 
is attained by the pair of strategies $\left(F_P(p_M),F_J(j_M)\right)$ satisfying:
\be\label{gp1_01_nocsi}
F_P(x)\sim k_p\mathbb{U}([\mu' \sigma_N^2,2v\mu'+\mu' \sigma_N^2])(x)+(1-k_p)\Delta_0(x),\nonumber
\ee 
\be\label{gp1_02_nocsi}
F_J(y)\sim \frac{2v}{2v+\sigma_N^2}k_j\mathbb{U}([0,2v])(y)+(1-\frac{2v}{2v+\sigma_N^2}k_j)\Delta_0(y),\nonumber
\ee 
where $\mathbb{U}([r,t])(\cdot)$ denotes the CDF of a uniform distribution
over the interval $[r,t]$, and $\Delta_0(\cdot)$ denotes the CDF of a Dirac distribution (i.e. a step function),
and the parameters $k_p,k_j\in [0,1]$ and $v\in [\max \{\mathcal{J}, \mathscr{J'}_M(\mathcal{P})/2 \}, \infty)$ are uniquely
determined from the following steps:

\begin{enumerate}
\item If 
\be
\mathcal{P}\geq \mu'\sigma_N^2+\frac{1}{2}\mu'\mathcal{J}\left[1+\sqrt{1+\frac{2\sigma_N^2}{\mathcal{J}}} \right],
\ee
then 
\be
v=\frac{\mathcal{P}-\mu'\sigma_N^2}{\mu'},
\ee
\be
k_p=1
\ee
and 
\be
k_j=\frac{\mu'\mathcal{J}(2\mathcal{P}-\mu'\sigma_N^2) }{2(\mathcal{P}-\mu'\sigma_N^2)^2}.
\ee

\item If
\be
\mathcal{P}<\mu'\sigma_N^2+\frac{1}{2}\mu'\mathcal{J}\left[1+\sqrt{1+\frac{2\sigma_N^2}{\mathcal{J}}} \right],
\ee
then
\be
v=\frac{1}{2}\mathcal{J}\left[1+\sqrt{1+\frac{2\sigma_N^2}{\mathcal{J}}} \right],
\ee
\be
k_p=\frac{2v\mathcal{P}}{\mu'(2v+\sigma_N^2)(2v-\mathcal{J})}
\ee
and 
\be
k_j=1.
\ee
\end{enumerate}
\end{thm}
\vspace*{4pt}
\begin{proof}
The proof follows directly from Theorem \ref{thm_gp1_1}  in Appendix \ref{app8},
by substituting $x=P_M$, $y=J_M$, $g(y)=\mathscr{P'}_M(y)$, $g^{-1}(x)=\mathscr{J'}_M(x)$,
$a=\mathcal{P}$ and $b=\mathcal{J}$. It is also interesting to note that the condition
$\int_0^{b}g(y)dy<\int_{g(b)}^{\infty}g^{-1}(x)dx$ is satisfied because $\mathscr{P'}_M(y)$
is unbounded.
\end{proof}
\vspace*{4pt}


\subsection{Numerical results}\label{ss4_3}

In this subsection we provide the numerical evaluation of our system's performance
when no channel state information is fed back by the receiver.
The parameters are identical to those used in the numerical evaluation of the previous sections.

\begin{figure}[h]
\centering
\includegraphics[scale=0.5]{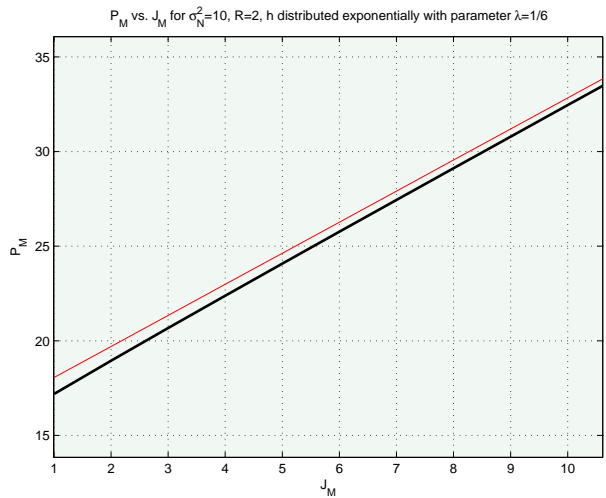}
\caption{$P_M$ vs. $J_M$ curve with and without CSI feedback when $R=2$, $\sigma_N^2=10$ and $h$ is distributed exponentially, with
parameter $\lambda=1/6$.}\label{PMJMcurve_nocsi}
\end{figure}

The new $\mathscr{P}_M(J_M)$ curve is given in Figure \ref{PMJMcurve_nocsi}. It can be seen that for a given jamming power
allocated to a frame, the transmitter power
required to ensure asymptotically error-free transmission over that frame is only slightly larger if no CSI is fed back
than when full CSI is available to all parties.

\begin{figure}[h]
\centering
\includegraphics[scale=0.5]{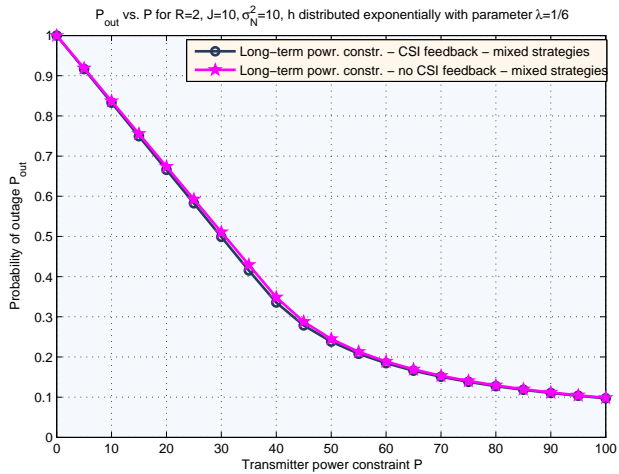}
\caption{Outage probability vs. transmitter power constraint $\mathcal{P}$ with and without CSI feedback when $\mathcal{J}=10$,
 $R=2$, $\sigma_N^2=10$ and $h$ is distributed exponentially, with
parameter $\lambda=1/6$. (Mixed strategies.)}\label{fig_numericalresults3_nocsi}
\end{figure}

\begin{figure}[h]
\centering
\includegraphics[scale=0.5]{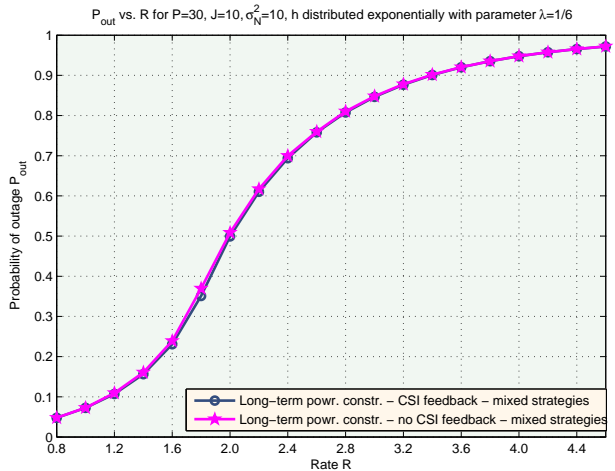}
\caption{Outage probability vs. rate with and without CSI feedback for $\mathcal{P}=30$,
$\mathcal{J}=10$, $\sigma_N^2=10$ and $h$ is distributed exponentially, with
parameter $\lambda=1/6$.(Mixed strategies.)}\label{fig_numericalresults4_nocsi}
\end{figure}

This observation explains the very small difference in achievable outage probabilities that can be observed in
Figures \ref{fig_numericalresults3_nocsi}  and \ref{fig_numericalresults4_nocsi}.

\section{Conclusions}\label{section5}

We have shown that for a high transmission rate $R$ the jammer could have enough power to keep the
ergodic capacity below $R$. In this scenario, if the transmitter imposes
average power constraints rather than peak power constraints, reliable communication is possible at the cost
of a non-zero probability of outage.

If both transmitter and jammer use average power constraints, their optimal strategies result as
solutions of a two-person zero-sum game. This game is played on two levels of power control.
The second level (power control within a frame) exhibits similar strategies for the pure
(maximin and minimax cases) and mixed strategies scenarios.
However in the pure strategies scenario, maximin and minimax first level power control (between frames) is generally
done differently, implying the non-existence of a Nash equilibrium.
A Nash equilibrium was derived for the mixed strategies scenario, placing the value of the objective function between
those of the minimax and maximin pure strategies solutions.

Although it may seem that the mixed strategies game makes more sense from a practical point of view,
the pure strategies minimax scenario may be a more appropriate model for the case when the jammer does not
attempt to jam unless it senses that the transmitter is on. In any circumstances, the minimax scenario
with pure strategies serves as a lower bound (the pessimistic approach) to the system's performance.

The feedback of CSI by the legitimate receiver is known to bring benefits (in terms of achievable transmission rate)
when nobody attempts to jam the transmission. However, for a fast fading AWGN channel, these improvements are shown
to be marginal \cite{goldsmith}.
We have shown that a similar conclusion holds (this time in terms of outage probability) for the case when the parties
that communicate over the fast fading AWGN channel are under attack from a jammer. The CSI fed back can easily
be intercepted by the jammer, which can then use this information to the transmitter's disadvantage. If one should
also take into account the loss of bandwidth and the complexity required for CSI feedback and processing, keeping
the transmitter (and jammer) ignorant of the channel coefficients may seem a better choice.

The same remark cannot be made for a parallel slow fading AWGN channel. It was shown in \cite{caire} that
when CSI is fed back and no jamming is present, the improvements in terms of probability of outage
are significant. In Part II of this paper \cite{myself4} we show that this conclusion also holds if we consider
the jamming scenario. In doing this we exploit the similarities that the parallel slow fading channel bears to
the fast fading channel, and develop new and even more interesting techniques to make up for the additional
complexity incurred by this new model.


\appendices

\section{Peak Power Constraints - Proof of Theorem \ref{thm_short_term}}\label{app1}

This proof follows the one described in the Appendix B of \cite{caire}.
The probability of outage can be written as:
\begin{gather}\label{ap11}
Pr(C(P(h),J(h))<R)=E[\chi_{\{  C(P(h),J(h))<R \}}],
\end{gather}
where $\chi_{\{\mathscr{A}\}}$ denotes the indicator function of the set $\mathscr{A}$.
Replacing the power allocations by the solutions of the game described by (\ref{game21}) and (\ref{game22}), we define
\begin{gather}
\chi^*=\chi_{\{ C(P^*(h),J^*(h))<R \}}.
\end{gather}

We next use the fact that the pair $(P^*(h), J^*(h))$ determines an equilibrium of the game (\ref{game21}), (\ref{game22}).
Thus, for any random power allocation $P(h)$ satisfying the power constraint, we can write:
\begin{gather}\label{ap12}
\chi^*\leq \chi_{\{C(P(h),J^*(h))<R \}} , \textrm{with probability 1}.
\end{gather}
Similarly, for any random $J(h)$, we have
\begin{gather}\label{ap13}
\chi^*\geq \chi_{\{C(P^*(h),J(h))<R \}} , \textrm{with probability 1}.
\end{gather}

Now pick some arbitrary power allocation functions $P_a(h)$ and $J_a(h)$, which satisfy the peak power constraints,
and set 
\begin{gather}
\widehat{P}(h)=(1-\chi^*)P^*(h)+\chi^*P_a(h),
\end{gather}
and
\begin{gather}
\widehat{J}(h)=(1-\chi^*)J_a(h)+\chi^*J^*(h),
\end{gather}
It is easy to see that $\expec_h\widehat{P}(h)\leq \mathcal{P}$ with probability $1$ ,
$\expec_h\widehat{J}(h)\leq \mathcal{J}$ with probability $1$, and moreover that
\begin{gather}
\chi^*=\chi_{\{C(\widehat{P}(h),\widehat{J}(h))<R \}}.
\end{gather}

Note that transmitter and jammer could pick $P_a(h)=0$ and $J_a(h)=0$ respectively,
but this strategy would not improve their performances (power cannot be saved),
since the only power constraints are set over frames.

Now, using (\ref{ap11}), (\ref{ap12}) and (\ref{ap13}), we get:
\begin{gather}
Pr(C(P(h),\widehat{J}(h))<R)\geq {}\nonumber\\
{}\geq Pr(C(\widehat{P}(h),\widehat{J}(h))<R)\geq{} \nonumber\\
{}\geq Pr(C(\widehat{P}(h),J(h))<R),
\end{gather}
which proves the existence of a Nash equilibrium of the original game.


\section{Average Power Constraints: Pure Strategies}

\subsection{Proof of Proposition \ref{prop_lt1}}\label{app6}

In proving the proposition, we take a contradictory approach.
It suffices to show that the situation $J(h)>0$ and $\lambda<x(h)/h$ cannot be part of the solution
of \emph{Problem 1}.

Assume that $J(h)>0$ and $\lambda<x(h)/h$ for $h$ in some set $\mathscr{S}\subset\mathbb{R}_+$.
If the jammer decreases the value of $J(h)$ on $\mathscr{S}$, two situations are possible.
In the first one, $J(h)$ is reduced to zero on $\mathscr{S}$, and the transmitter is still ''absent''.
This happens if $\sigma_N^2>\lambda h$. In this case, modifying the value of $J(h)$
has no impact upon the value of $\lambda$, and hence neither upon the outcome.

In the second case $J(h)$ is reduced to some positive value $J'(h)$, such that the transmitter decides
to be ''non-absent'' over $\mathscr{S}$.
This happens if $J'(h)+\sigma_N^2=\lambda' h$.
Note that the value of $\lambda$ might be changed to some $\lambda'$.
However, as we shall see briefly, if we consider $J'(h)$ that satisfies $J'(h)+\sigma_N^2=\lambda' h$,
then we have $\lambda'=\lambda$.

To prove this, let $\lambda$ be given by (\ref{exprlambda}), and assume that
$\lambda-x(h)/h\geq 0$ for $h\in \mathscr{M'}$, and $\lambda-x(h)/h<0$ for
$h\in\mathscr{S}$.
Now modify $x(h)$ by decreasing $J(h)$ as above.
We have
\begin{eqnarray}\label{lamb'rel}
\lambda'=c^{\frac{1}{\mathfrak{m}(\mathscr{M'}\bigcup \mathscr{S})}}\left\{ \exp\left[
\expec_{h\in \mathscr{M'}\bigcup \mathscr{S}}\left(
\log \frac{x(h)}{h}\right)\right]\right\}^{\frac{1}{\mathfrak{m}(\mathscr{M'}\bigcup \mathscr{S})}}\nonumber\\
{}=\frac{x(h)}{h}, ~\textrm{for}~h\in \mathscr{S}.
\end{eqnarray}
Note that for $h\in \mathscr{S}$ we have $\frac{x(h)}{h}=\lambda'$, so
\begin{gather}\label{expectrel}
\expec_{h\in \mathscr{S}}\log\frac{x(h)}{h}=\log\lambda'\mathfrak{m}(\mathscr{S}).
\end{gather}

Taking logarithm of (\ref{lamb'rel}):
\begin{gather}\label{sssssssss}
\frac{1}{\mathfrak{m}(\mathscr{M'})+\mathfrak{m}(\mathscr{S})}\Bigg[
\log c+\expec_{h\in \mathscr{M'}}\left(\log \frac{x(h)}{h}\right)+\nonumber\\
+\expec_{h\in \mathscr{S}}\left(\log \frac{x(h)}{h}\right)
 \Bigg]=\log \frac{x(h)}{h}, ~\textrm{for}~h\in \mathscr{S},
\end{gather}
and noting that the left hand side of (\ref{sssssssss}) is independent of the actual realizations of $h$,
we can compute the expectation over $h\in \mathscr{S}$, and get:
\begin{gather}
\frac{\mathfrak{m}(\mathscr{S})}{\mathfrak{m}(\mathscr{M'})+\mathfrak{m}(\mathscr{S})}\left[
R+\expec_{h\in \mathscr{M'}}\left(\log \frac{x(h)}{h}\right)
\right]=\nonumber\\
=\frac{\mathfrak{m}(\mathscr{M'})}{\mathfrak{m}(\mathscr{M'})+\mathfrak{m}(\mathscr{S})}
\expec_{h\in \mathscr{S}}\left(\log \frac{x(h)}{h}\right).
\end{gather}
Using (\ref{expectrel}), this leads to
\begin{gather}
\log\lambda=\frac{1}{\mathfrak{m}(\mathscr{M'})}\left[
R+\expec_{h\in \mathscr{M'}}\left(\log \frac{x(h)}{h}\right)
\right]=\nonumber\\
=\frac{1}{\mathfrak{m}(\mathscr{S})}
\expec_{h\in \mathscr{S}}\left(\log \frac{x(h)}{h}\right)=\log\lambda'.
\end{gather}

Therefore the outcome is maintained because,
although ``non-absent'', the transmitter still invests zero power on $\mathscr{S}$.

Hence if such a situation where the jammer transmits on a set of channel coefficient values over which the transmitter
is ``absent'' occurs, the jammer can save power and maintain the same outcome.
Meanwhile the new set over which jammer transmits becomes a subset of the new set over which 
the transmitter is ``non-absent''.


\subsection{Proof of Proposition \ref{propmu}}\label{app7}

We already know that the optimal $x(h)$ is a continuous function of $h\in\mathscr{M''}$ if
$\mathscr{M'}$ and $\mathscr{M''}$ are fixed.

The following lemma shows that under this scenario the optimal $x(h)$ is also unique.
\vspace*{4pt}
\begin{lemma}\label{lemma1}
For fixed $\mathscr{M'}$ and $\mathscr{M''}$, the KKT conditions (\ref{kkt0})--(\ref{kkt111}) admit a unique solution.
\end{lemma}
\vspace*{4pt}
\begin{proof}
Consider $\mathscr{M'}$ and $\mathscr{M''}$ to be fixed.
The constant $\mu$ resulting from (\ref{kkt0})--(\ref{kkt111}) can be computed
as in (\ref{mu}).
This implies that $J_M(\mu)$ is a strictly decreasing function, hence an injection.

Thus, for a given $J_M$ there exists a unique corresponding value of $\mu$, and
since $x(h)$ is a deterministic function of $\mu$, a unique solution
$x(h)$.
\end{proof}
\vspace*{4pt}

Suppose the jammer's optimal power distribution $x^*(h)$ is not continuous over the whole $\mathbb{R}_+$.

Note that an optimal power distribution $x^*(h)$ obtained for fixed
$\mathscr{M'}$ and $\mathscr{M''}$ can only be a globally optimal solution (i.e. over
all possible choices of $\mathscr{M'}$ and $\mathscr{M''}$), if by keeping
the same $\mathscr{M'}$ and extending $\mathscr{M''}$ to a set
$\mathscr{M''}_n$ that contains a discontinuity point
, the new optimal strategy is either the same as
$x^*(h)$, or violates the constraint $x(h)\geq \sigma_N^2$.
But an optimal strategy has to be continuous over $\mathscr{M''}_n$, and hence
the constraint $x(h)\geq \sigma_N^2$ has to be violated on the left-most side of $\mathscr{M''}_n$
(according to (\ref{relxnx})).

Also note that if under the optimal strategy the jammer allocates some power $J_x$ over a 
set $\mathscr{M_x}\subset \mathbb{R}_+$, then the distribution of $J_x$ over $\mathscr{M_x}$
should be done optimally, according to (\ref{relxnx}), (\ref{mu}).
This implies that by extending the set $\mathscr{M_x}$ by a set $\mathscr{N}$ disjoint from
$\mathscr{M''}$, and re-allocating $J_x$ over $\mathscr{M}_x \bigcup \mathscr{N}$, the constraint
$x(h)\geq \sigma_N^2$ will be violated on the left-most side of $\mathscr{M}_x \bigcup \mathscr{N}$.

The arguments above imply the following:
\begin{enumerate}
\item The optimal jamming power allocation should be such that $x(h)=\sigma_N^2$ on the left-most point of $\mathscr{M''}$:
otherwise extend $\mathscr{M''}$ by an arbitrarily small set to the left and increase $J_M$ until $x(h)=\sigma_N^2$ on the left-most
point of the new set $\mathscr{M''}_n$; by continuity of $x(h)$, the left-most point of $\mathscr{M''}$ should be arbitrarily close
to $\sigma_N^2$.
\item The optimal jamming power allocation should be such that $\mathscr{M''}=[h^*,\infty)$: otherwise
take a subset $\mathscr{M}_x\subset \mathscr{M''}$, such that there exists a set $\mathscr{N}$ situated
to the right of $\mathscr{M}_x$, and denote by $J_x$ the jamming power originally allocated to $\mathscr{M}_x$.
By re-allocating $J_x$ over $\mathscr{M}_x \bigcup \mathscr{N}$, the constraint
$x(h)\geq \sigma_N^2$ will be violated on the left-most side of $\mathscr{M}_x$.
If $\mathscr{N}$ is picked of arbitrarily small $\mathfrak{m}$-measure, by the previous arguments
we should have $x(h)$ arbitrarily close to $\sigma_N^2$ at the left-most point of $\mathscr{M}_x$.
But since $\mathscr{M}_x$ is arbitrary, this yields the contradiction that $x(h)=\sigma_N^2$ for
any $h$ to the left of  $\mathscr{N}$.
\end{enumerate}
This proves Proposition \ref{propmu}.

Note that if $\mu=0$, then $P(h)=0$ over $\mathscr{M''}$, and since $x(h)/h$ is decreasing over the whole $\mathbb{R}_+$,
and $\mathscr{M''}=[h^*,\infty)$, this implies that the transmitter does not transmit at all.
However, this strategy does not achieve an ergodic capacity larger than the rate $R$, and hence it results in a contradiction.


\subsection{Proof of Proposition \ref{propconcave}}\label{app5}

Recall Proposition \ref{propconcave}:
\emph{Under the optimal maximin second level power control strategies,
the ``required'' transmitter power $P_M$ over a frame is a strictly increasing, unbounded and concave function of the power
$J_M$ that the jammer invests in that frame.}

The fact that $\mathscr{P}_M(J_M)$ is strictly increasing follows from Proposition \ref{propuniq} and Proposition
\ref{circ_pr_prop1}. If $J_{M,1}<J_{M,2}$ existed such that $P_M(J_{M,1})=P_M(J_{M,2})$, then when the jammer's
power constraint is $J_{M,2}$, \emph{Problem 1} would either have two different solutions, or the solution would
satisfy the constraint with strict inequality.

If $J_M \to \infty$ then (\ref{relxnx}) implies that $J(h)\to \infty$ for any $h$. If $P_M$ was finite, this
would imply $C(P(h),J(h))\to 0$, which violates the constraints of \emph{Problem 1}.
Hence $\mathscr{J}_M(P_M)$ has to be unbounded.

In proving concavity of the $\mathscr{P}_M(J_M)$ function for the case when the channel coefficient $h$ belongs to
a continuous alphabet, we first show that the solution of the discretized problem (i.e. when $h$ belongs
to a discrete alphabet, obtained by some discretization of the original continuous alphabet)
is unique and converges point-wise to the solution of the continuous problem as the discrete
alphabet converges to the original continuous alphabet.

This approach also serves the purpose of legitimizing numerical evaluations.

Next, we prove that for the discretized problem $\mathscr{P}_M(J_M)$ is concave.
Finally, we show that point-wise convergence of a sequence of concave functions is enough for the concavity of its limit function.

Consider the uniformly spaced discretization $q\mathbb{Z_+}$ of the interval $[0,\infty)$, and a p.m.f. of the channel coefficient
$h\in q\mathbb{Z_+}$ that converges to the original p.d.f. as $q$ goes to zero.

The maximin second level power allocation problem can still be written as in (\ref{probl1}), even though
the integrals representing the expectations can now be written as sums.
Moreover, Propositions \ref{prop_lt1}--\ref{propmu} and relations (\ref{Pn_first_expr})--(\ref{mu}) hold with the only
modification that the term ``continuous'' should be crossed out.

The second level power allocation solution for the discretized maximin problem is completely determined by
the triple $(\mathscr{M'}, \mathscr{M''}, \mu)$, or equivalently by $(h^0, h^*, \mu)$.
Instead of (\ref{sys1})--(\ref{sys3}) we can now write

\begin{eqnarray}\label{sys10d}
\frac{\sigma_N^2}{h^0}\leq\lambda < \frac{\sigma_N^2}{h^0-q},
\end{eqnarray}
\begin{eqnarray}\label{sys20d}
\sigma_N^2\frac{1+\mu h^*}{h^*}\leq\lambda < \sigma_N^2\frac{1+\mu (h^*-q)}{h^*-q},
\end{eqnarray}
\begin{eqnarray}\label{sys30d}
R=\sum_{h_0}^{h^*-q}\log\left(\frac{\lambda h}{\sigma_N^2}\right)p(h)-{}\nonumber\\
{}-\sum_{h^*}^{\infty}\log\left(\frac{1}{1+\mu h}\right)p(h),
\end{eqnarray}
or equivalently
\begin{eqnarray}\label{sys1d}
Q_U\left[\frac{h^*-q}{1-\mu (h^*-q)}\right]\leq h^0 \leq Q_D\left[\frac{h^*}{1-\mu h^*}+q\right],
\end{eqnarray}
\begin{eqnarray}\label{sys2d}
\sum_{h=h^*}^{\infty}\left[\frac{\frac{h}{1+\mu h}}{\frac{h^*}{1+\mu h^*}}-1\right]p(h)\leq\frac{J_M}{\sigma_N^2} \leq \nonumber\\
\leq\sum_{h=h^*}^{\infty}\left[\frac{\frac{h}{1+\mu h}}{\frac{h^*-q}{1+\mu (h^*-q)}}-1\right]p(h),
\end{eqnarray}
\begin{eqnarray}\label{sys3d}
\sum_{h=Q_D\left[\frac{h^*}{1+\mu h^*}+q\right]}^{h^*-q}\log\left(h\frac{1+\mu h^*}{h^*}\right)p(h)-{}\nonumber\\
{}-\sum_{h^*}^{\infty}\log\left(\frac{1}{1+\mu h}\right)p(h)\leq R \leq{}\nonumber\\
{}\leq\sum_{h=Q_U\left[\frac{h^*-q}{1+\mu (h^*-q)}\right]}^{h^*-q}\log\left(h\frac{1+\mu(h^*-q)}{h^*-q}\right)p(h)-{}\nonumber\\
{}-\sum_{h^*}^{\infty}\log\left(\frac{1}{1+\mu h}\right)p(h),
\end{eqnarray}
where $Q_D[h]$ denotes the largest element of $q\mathbb{Z_+}$ that is less than $h$ and
$Q_U[h]$ denotes the smallest element of $q\mathbb{Z_+}$ that is larger than $h$.

\vspace*{4pt} 
\begin{lemma}\label{lemmaunique}
For a given $J_M$ the solution of the discretized maximin second level power allocation problem is unique.
\end{lemma}
\vspace*{4pt} 
\begin{proof}
It is straightforward to show that for fixed $h^*$ the left-most and the right-most terms of inequality (\ref{sys2d})
(which upper-bound and lower-bound $J_M/ \sigma_N^2$)
are strictly decreasing functions of $\mu$, and similarly the left-most and the right-most terms of inequality (\ref{sys3d})
are strictly increasing functions of $\mu$.

Note that
\be
\sum_{h=h^*}^{\infty}\left[\frac{\frac{h}{1+\mu h}}{\frac{h^*}{1+\mu h^*}}-1\right]p(h)=
\sum_{h=h^*+q}^{\infty}\left[\frac{\frac{h}{1+\mu h}}{\frac{h^*}{1+\mu h^*}}-1\right]p(h),
\ee
\be
Q_D\left[\frac{h^*-q}{1+\mu (h^*-q)}+q\right]=Q_U\left[\frac{h^*-q}{1+\mu (h^*-q)}\right],
\ee
and
\be
\sum_{h=Q_D\left[\frac{h^*}{1+\mu h^*}+q\right]}^{h^*-q}\log\left(h\frac{1+\mu h^*}{h^*}\right)p(h)-{}\nonumber\\
{}-\sum_{h^*}^{\infty}\log\left(\frac{1}{1+\mu h}\right)p(h)={}\nonumber\\
{}=\sum_{h=Q_D\left[\frac{h^*}{1+\mu h^*}+q\right]}^{h^*}\log\left(h\frac{1+\mu h^*}{h^*}\right)p(h)-{}\nonumber\\
{}-\sum_{h^*+q}^{\infty}\log\left(\frac{1}{1+\mu h}\right)p(h).
\ee
These arguments imply that by keeping $\mu$ constant and replacing $h^*$ by $h^*-q$ in both first terms of
(\ref{sys2d}) and (\ref{sys3d}), we get exactly the last terms of (\ref{sys2d}) and (\ref{sys3d}), respectively.
Thus, if ($h^*$,$\mu$) satisfy both (\ref{sys2d}) and (\ref{sys3d}), then decreasing $h^*$ (by more than one step) and maintaining the same
$\mu$ violates both (\ref{sys2d}) and (\ref{sys3d}).
In order for (\ref{sys2d}) to still hold, $\mu$ should be increased, while in order for (\ref{sys3d}) to still hold,
$\mu$ should be decreased.
But once $h^*$ and $\mu$ are given, $\lambda$ and hence $h^0$ are
uniquely determined. Therefore there cannot exist more than one solution to the discretized problem.
\end{proof}
\vspace*{4pt} 
The following lemma deals with the convergence of this solution as $q\to 0$.
\vspace*{4pt} 
\begin{lemma}\label{lammaconverg}
For a given $J_M$, the solution of the discretized maximin second level power allocation problem converges to
the solution of the continuous problem as $q\to 0$.
\end{lemma}
\vspace*{4pt} 
\begin{proof}
This follows by noticing that as $q\to 0$ (\ref{sys10d})--(\ref{sys30d}) become arbitrarily close to (\ref{sys1})--(\ref{sys3}),
and the sums involved in the expectations converge to integrals (by the definition of the Riemann integral).
\end{proof}

Next we prove that for the discretized problem, the resulting $\mathscr{P}_M(J_M)$ function is concave.
We first show in Lemma \ref{propapp51} that
the optimal jammer strategy $\{x^*(h)\}_{h=0}^{\infty}$ is a continuous function of
the given jamming power $J_M$.
Lemma \ref{propapp52} proves that $P_M(\{x(h)\})$ is continuous and has
continuous first order derivatives.
This implies that $\mathscr{P}_M(J_M)$ is in fact continuous and has a continuous first order derivative.
Finally, Lemma \ref{propapp53} shows that for any fixed $M'$ and $M''$ the function $\mathscr{P}_M(J_M)$ is concave.

\vspace*{4pt}
\begin{lemma}\label{propapp51}
The optimal jammer power allocation $\{x^*(h)\}_{h\in q\mathbb{Z_+}}$ within a frame
is a continuous increasing function of the given jamming power $J_M$ invested over that frame.
\end{lemma}
\vspace*{4pt}
\begin{proof}
It is clear that $x(h)$ is continuous and increasing as a function of $J_M$ if $h^*$ and $h^0$ are fixed.
At any point where either $h^*$ or $h^0$ change as a result of a change in $J_M$, the optimal jamming strategy
$\{x^*(h)\}_{h\in q\mathbb{Z_+}}$ maintains continuity as a result of the uniqueness of the solution (Lemma \ref{lemmaunique}).
\end{proof}
\vspace*{4pt}

\begin{lemma}\label{propapp52}
Both $P_M(\{x(h)\})$ and the derivatives $\frac{dP_M}{dx(h)}$, for $h\in q\mathbb{Z_+}$ are continuous
functions of $\{x(h)\}_{h\in q\mathbb{Z_+}}$.
\end{lemma}
\vspace*{4pt}
\begin{proof}
Consider any two points $\{x_1(h)\}_{h\in q\mathbb{Z_+}}$ and $\{x_2(h)\}_{h\in q\mathbb{Z_+}}$ and any
trajectory $\mathfrak{T}$ that connects them.

Without loss of generality, assume that the channel coefficients are always indexed
in decreasing order of the quantities $\frac{x(h)}{h}$.

For a given vector $\{x(h)\}_{h\in q\mathbb{Z_+}}$, the required transmitter power is
\begin{eqnarray}\label{relPM}
P_M=\lambda\sum_{h\in\mathscr{M'}}p(h)-\sum_{h\in\mathscr{M'}}\frac{x(h)}{h}p(h),
\end{eqnarray}
while the derivatives are given by
\begin{eqnarray}\label{relderPM}
\frac{dP_M}{dx(h)}=\left[\frac{\lambda}{x(h)}-\frac{1}{h}\right]p(h)
\end{eqnarray}
for $h\in\mathscr{M'}$, with $\lambda$ given by
\begin{eqnarray}\label{relPM1}
\lambda(\mathscr{M'})=\left[c\prod_{h\in\mathscr{M'}}\left(\frac{x(h)}{h}\right)^{p(h)}\right]^{\frac{1}{\sum_{h\in\mathscr{M'}}p(h)}}.
\end{eqnarray}

Note that $\mathscr{M'}$ depends upon the choice of $\{x(h)\}$.
For fixed $\mathscr{M'}$, the continuity and differentiability of $P_M(\{x(h)\})$ are obvious.
Thus, it suffices to show that these properties also hold in a point of $\mathfrak{T}$ where $\mathscr{M'}$ changes.

If we can show continuity and differentiability when $\mathscr{M'}$ is increased by including
one channel coefficient $h_0$, then larger variations of $\mathscr{M'}$
can be treated as multiple changes by one channel coefficient, and continuity still holds.

Let $\{x_k(h)\}_{h\in q\mathbb{Z_+}}$ be a point of $\mathfrak{T}$ where
the transmitter increases the number of frames over which it transmits as above,
and denote by $\mathfrak{T_1}$ the part of the trajectory $\mathfrak{T}$ that is between
$\{x_1(h)\}$ and $\{x_k(h)\}$, and $\mathfrak{T_2}=\mathfrak{T}\setminus \mathfrak{T_1}$.

Since $P(h_0)=0$ (i.e. $\lambda=\frac{x(h_0)}{h_0}$),
we have $\lambda(\mathscr{M'})=\lambda(\mathscr{M'}\bigcup \{h_0\})$, because
they both satisfy
\begin{gather}
\sum_{h\in \mathscr{M'}}\left[\lambda-\frac{x(h)}{h} \right]p(h)=P_M.
\end{gather}

Define the ``left'' and ``right'' limits  $P_M(\{x_k(h)\}-)$ and $P_M(\{x_k(h)\}+)$ as:
\begin{gather}\label{leftrightlimit1}
P_M(\{x_k(h)\}-)=\lim_{\substack{\{x(h)\}\to \{x_k(h)\}\\ \{x(h)\}\in
\mathfrak{T_1}}} P_M(\{x(h)\}),
\end{gather}
\begin{gather}\label{leftrightlimit2}
P_M(\{x_k(h)\}+)=\lim_{\substack{\{x(h)\}\to \{x_k(h)\}\\ \{x(h)\}\in
\mathfrak{T_2}}} P_M(\{x(h)\}).
\end{gather}
We can now write: 
\begin{gather}
P_M(\{x(h)\}+)={}\nonumber\\
{}=\lambda\sum_{h\in\mathscr{M'}\bigcup \{h_0\}}p(h)-\sum_{h\in\mathscr{M'}\bigcup \{h_0\}}\frac{x(h)}{h}p(h)={}\nonumber\\
{}=\lambda\sum_{h\in\mathscr{M'}}p(h)-\sum_{h\in\mathscr{M'}}\frac{x(h)}{h}p(h)+{}\nonumber\\
{}+\lambda p(h_0)-\frac{x(h_0)}{h_0}p(h_0)=P_M(\{x(h)\}-)
\end{gather}
where the last equality follows since $\lambda=\frac{x(h_0)}{h_0}$.
This proves continuity.

Similar arguments can be used to show the continuity of the derivatives in (\ref{relderPM}).
\end{proof}
\vspace*{4pt}

\begin{lemma}\label{propapp53}
In the discretized case, for fixed $h^0$ and $h^*$, the function $\mathscr{P}_M(J_M)$ is concave.
\end{lemma}
\vspace*{4pt}
\begin{proof}

Write (\ref{mu}) explicitly for the discretized problem:
\begin{eqnarray}\label{reljm}
MJ_M+\sigma_N^2\sum_{h=h^*}^{\infty}p(h)=\left[c\prod_{h=h^*}^{\infty}\left(\frac{1}{1+\mu h}\right)^{p(h)}\right.\cdot\nonumber\\
\left.\cdot\prod_{h=h^0}^{h^*-q}\left(\frac{\sigma_N^2}{h}\right)^{p(h)}
\right]^{\frac{1}{\sum_{h=h^0}^{h^*-q}p(h)}}
\sum_{h=h^*}^{\infty}\frac{h}{1+\mu h}p(h),
\end{eqnarray}
and denote
\begin{eqnarray}\label{relgmu}
g(\mu)=\prod_{h=h^*}^{\infty}\left(\frac{1}{1+\mu h}\right)^{\frac{p(h)}{\sum_{h=h^0}^{h^*-q}p(h)}}\cdot
\sum_{h=h^*}^{\infty}\frac{h}{1+\mu h}p(h)
\end{eqnarray}
Note that for fixed $h^0$ and $h^*$, $J_M$ is a linear function of $g$.

From (\ref{Pn_first_expr}), (\ref{exprlambda}) and (\ref{relxnx})
a similar relation can be found for the required transmitter power $P_M$:
\begin{eqnarray}\label{relpm}
MP_M+\sum_{h=h^0}^{h^*-q}\frac{\sigma_N^2}{h_m}p(h)=\left[c\prod_{h=h^*}^{\infty}\left(\frac{1}{1+\mu h}\right)^{p(h)}\right.\cdot\nonumber\\
\left.\cdot\prod_{h=h^0}^{h^*-q}\left(\frac{\sigma_N^2}{h}\right)^{p(h)}
\right]^{\frac{1}{\sum_{h=h^0}^{h^*-q}p(h)}}\cdot\nonumber\\
\cdot\left[\sum_{h=h^0}^{h^*-q}p(h)-\sum_{h=h^*}^{\infty}\frac{1}{1+\mu h}p(h)\right].
\end{eqnarray}
Denote
\begin{eqnarray}\label{relfmu}
f(\mu)=\prod_{h=h^*}^{\infty}\left(\frac{1}{1+\mu h}\right)^{\frac{p(h)}{\sum_{h=h^0}^{h^*-q}p(h)}}\cdot\nonumber\\
\cdot\left[\sum_{h=h^0}^{\infty}p(h)-\sum_{h=h^*}^{\infty}\frac{1}{1+\mu h}p(h)\right],
\end{eqnarray}
and note that for fixed $h^0$ and $h^*$, $P_M$ is a linear function of $f$.

It suffices to show that $f(g)$ is concave.
For this purpose, the derivative $\frac{df}{dg}=\frac{df}{d\mu}(\frac{d\mu}{dg})^{-1}$ should be
a decreasing function of $g$, and hence an increasing function of $\mu$.

Computing the derivatives from (\ref{relgmu}) and (\ref{relfmu}) we obtain

\begin{eqnarray}\label{relg}
\frac{df}{dg}=\frac{\frac{df}{d\mu}}{\frac{dg}{d\mu}}=
\frac{\frac{1}{\sum_{h=h^0}^{h^*-q}p(h)}\left(\sum_{h=h^0}^{\infty}p(h)-\sum_{h=h^*}^{\infty}\frac{1}{1+\mu h}p(h)\right)
-\frac{\sum_{h=h^*}^{\infty}\frac{h}{(1+\mu h)^2}p(h)}{\sum_{h=h^*}^{\infty}\frac{h}{1+\mu h}p(h)}}
{\frac{1}{\sum_{h=h^0}^{h^*-q}p(h)}\sum_{h=h^*}^{\infty}\frac{h}{(1+\mu h)^2}p(h)+
\frac{\sum_{h=h^*}^{\infty}\frac{h^2}{(1+\mu h)^2}p(h)}{\sum_{h=h^*}^{\infty}\frac{h}{1+\mu h}p(h)}}
\end{eqnarray}

Looking at the right hand side of (\ref{relg}) (the ``large fraction''), we notice that the
first term in the numerator increases with $\mu$.
For the second term in the numerator, it is clear that as $\mu$ increases, its numerator decreases
faster than its denominator. This implies that the whole numerator of the ``large fraction'' is an increasing function of $\mu$.
Similarly, the first term in the denominator is clearly a decreasing function of $\mu$.
The only thing left is the second term of the denominator.
It is straightforward to show that its derivative with respect to $\mu$
can be written as
\be\label{jvouscblajkghilu}
\frac{d}{d\mu}\frac{\sum_{h=h^*}^{\infty}\frac{h^2}{(1+\mu h)^2}p(h)}{\sum_{h=h^*}^{\infty}\frac{h}{1+\mu h}p(h)}
=\frac{1}{\left[\sum_{h=h^*}^{\infty}\frac{h}{1+\mu h}p(h)\right]^2}\cdot\nonumber\\
\cdot\Bigg\{ \left[\sum_{h=h^*}^{\infty}\frac{h^2}{(1+\mu h)^2}p(h)\right]^2-\sum_{h=h^*}^{\infty}\frac{h^3}{(1+\mu h)^3}p(h)\cdot\nonumber\\
\cdot \sum_{h=h^*}^{\infty}\frac{h}{(1+\mu h)}p(h)\Bigg\}
\ee

If we consider the fact that for any two real numbers $a$ and $b$ we have
\be
(a^2+b^2)^2-(a+b)(a^3+b^3)=-ab(a-b)^2
\ee 
and the summations in (\ref{jvouscblajkghilu}) are positive, it is easy to see that the second term of the denominator
of the ``large fraction'' is decreasing with $\mu$. 
Hence overall the derivative in (\ref{relg}) increases with $\mu$.
\end{proof}
\vspace*{4pt}

\begin{lemma}\label{lammaconcavelimit}
The limit of a point-wise convergent sequence of concave functions is concave.
\end{lemma}
\vspace*{4pt} 
\begin{proof}
Denote the sequence by $(f_n(x))_{n=1}^\infty$ and its limit by $f(x)$.
Point-wise convergence implies that for any $x$ and $\forall \epsilon>0$, $\exists N(x)$ such that
$|f(x)-f_n(x)|<\epsilon$, $\forall n\geq N(x)$.

Take two arbitrary points $x$ and $y$, and pick some arbitrary $\alpha\in [0,1]$.
Denote $N=\max\{N(x),N(y), N(\alpha x+(1-\alpha)y)\}$. 
Then for $n\geq N$ and any $\epsilon>0$ we have
\begin{gather}
f(\alpha x+(1-\alpha)y)>f_n(\alpha x+(1-\alpha)y)-\epsilon \geq{}\nonumber\\
{}\geq \alpha f_n(x)+(1-\alpha)f_n(y)-\epsilon>{}\nonumber\\
{}> \alpha f(x)+(1-\alpha)f(y)-2\epsilon,
\end{gather}
where the second inequality follows from the concavity of $f_n$.
This implies that $f$ is also concave.
\end{proof}


\subsection{On a special kind of duality}\label{app4}

Take $x,y \in L^2[\mathbb{R}]$ and define the order relation $x>y$ if and only if $x(t)>y(t)~\forall
t\in \mathbb{R}$. 
Consider the continuous real functions $f(x)$, $g(y)$ and $h(x,y)$ over $L^2[\mathbb{R}]$, such that
$f$ is a strictly increasing function of $x$, $g$ is a strictly increasing function of $y$,
and $h$ is a strictly increasing function of $x$ for fixed $y$ and
a strictly decreasing function of $y$ for fixed $x$.

Define the following minimax and maximin problems:
\be \label{genpr1}
\max_{y\geq0}\left[ \min_{x\geq0} f(x) ~\textrm{s.t.}~ h(x,y)\geq H \right]
\textrm{s.t.} g(y)\leq G,
\ee

\be \label{genpr2}
\max_{x\geq0}\left[ \min_{y\geq0} g(y) ~\textrm{s.t.}~ h(x,y)\leq H \right]
\textrm{s.t.} f(x)\leq F,
\ee

\be \label{genpr3}
\min_{y\geq0}\left[ \max_{x\geq0} h(x,y) ~\textrm{s.t.}~ f(x)\leq F \right]
\textrm{s.t.} g(y)\leq G.
\ee

The following result is important in the proof of Theorem \ref{theoremgeneral} below.
\vspace*{4pt}
\begin{prop}\label{circ_pr_prop1}
For any of the three problems above, the optimal solution satisfies both constraints with equality.
\end{prop}
\vspace*{4pt}
\begin{proof}
Take problem (\ref{genpr1}).
Let $(x_1,y_1)$ be a solution such that $f(x_1)=F$,
and assume that $h(x_1,y_1)>H$.
Since $h$ is a continuous, strictly increasing function of $x$ for a fixed $y$,
we can find $x_n<x_1$ such that $h(x_n,y_1)=H$.
But then $f(x_n)<f(x_1)$, which means that there exists a better value of
$x$ if $y=y_1$, and hence that $(x_1,y_1)$ is not a solution.

Therefore, the first constraint has to be satisfied with equality.

Now assume that $g(y_1)<G$. Then we can find $y_0>y_1$, such that $g(y_0)=G$.
However, since $h(x_1,y_1)=H$, we have $h(x_1,y_0)<H$.
In order for the first constraint to be satisfied, we need to replace $x_1$ by some other value
$x_0$. We prove next that the value $x_0$ resulting from this modification will be such that
$f(x_0)>f(x_1)$, which makes the pair $(x_1,y_1)$ suboptimal, thus
contradicting the hypothesis that it is a solution, and proving that the second constraint should hold with equality.

Assume that the value of $x_0$ is such that 
\begin{gather}\label{contra}
f(x_0)=F_0\leq F.
\end{gather}

Then, replacing $y_0$ by $y_1$, we have that $(x_0,y_1)$ is either a second solution of Problem 1
(if the inequality in (\ref{contra}) holds with equality), or a better choice (if the inequality in (\ref{contra})
holds with strict inequality).
We can readily dismiss the latter case, since $(x_1,y_1)$ was assumed to be an optimal solution.
For the former case, $h$ is a strictly decreasing function of $y$, thus $h(x_0,y_1)>R$,
which contradicts the first part of this proof.
The same arguments work for the problem in (\ref{genpr2}).

Take problem (\ref{genpr3}), and denote by $(x_3,y_3)$ one of its optimal solutions.
If $g(y_3)<G$, we can increase $y$ up to a value $y_m$ such that $g(y_m)=G$.
But in turn, this yields $h(x_3,y_m)<h(x_3,y_3)$, making $y_3$ suboptimal.
Therefore, the first constraint has to hold with equality.

Similarly, if $f(x_3)<F$, we can increase $x$ up to a value $x_m$ such that $f(f_m)=F$,
yielding $h(x_m,y_3)>h(x_3,y_3)$, and thus resulting in a contradiction.
Thus the second constraint also holds with equality. 
\end{proof}

The main result of this section is the following theorem, which introduces a special kind
of duality between the three problems in (\ref{genpr1}), (\ref{genpr2}) and (\ref{genpr3}).
\vspace*{4pt}
\begin{thm}\label{theoremgeneral}
(I) Choose any real values for $G$ and $H$.
Take problem (\ref{genpr1}) under these constraints and let the pair $(x_1,y_1)$ denote one of its optimal solutions,
yielding a value of the objective function $f(x_1)=F_1$. If we set the value of the corresponding
constraints in problems (\ref{genpr2}) and (\ref{genpr3}) to $F=F_1$, then the values of the objective
functions of problems (\ref{genpr2}) and (\ref{genpr3}) under their optimal solutions are
$g(y)=G$ and $h(x,y)=H$, respectively. Moreover, $(x_1,y_1)$ is also an optimal solution of all problems.

(II) Choose any real values for $F$ and $H$.
Take problem (\ref{genpr2}) under these constraints and let the pair $(x_2,y_2)$ denote one of its optimal solutions,
yielding a value of the objective function $g(y_2)=G_2$. If we set the value of the corresponding
constraints in problems (\ref{genpr1}) and (\ref{genpr3}) to $G=G_2$, then the values of the objective
functions of problems (\ref{genpr1}) and (\ref{genpr3}) under their optimal solutions are
$f(x)=F$ and $h(x,y)=H$, respectively. Moreover, $(x_2,y_2)$ is an optimal solution of all problems.
 
(III) Choose any real values for $F$ and $G$.
Take problem (\ref{genpr3}) under these constraints and let the pair $(x_3,y_3)$ denote one of its optimal solutions,
yielding a value of the objective function $h(x_3,y_3)=H_3$. If we set the value of the corresponding
constraints in problems (\ref{genpr1}) and (\ref{genpr2}) to $H=H_3$, then the values of the objective
functions of problems (\ref{genpr1}) and (\ref{genpr2}) under their optimal solutions are
$f(x)=F$ and $g(y)=G$, respectively. Moreover, $(x_3,y_3)$ is an optimal solution of all problems.
\end{thm}
\vspace*{4pt}
\begin{proof}
(I) Take problem (\ref{genpr1}) and let
$(x_1,y_1)$ be an optimal solution, such that  $f(x_1)=F$.
We need to show that $(x_1,y_1)$ is also an optimal solution of problems (\ref{genpr2}) and (\ref{genpr3}).

Since $x_1$ and $y_1$ form a solution of problem (\ref{genpr1}), by Proposition \ref{circ_pr_prop1}, they satisfy the first constraint
in problem (\ref{genpr1}) with equality, and so they also satisfy the first constraint in problem (\ref{genpr2}).
Furthermore, since the second constraint of problem (\ref{genpr2}) reads $f(x)\leq F$, we note that $x_1$ and $y_1$ are
in the feasible set.
If we evaluate the cost function at this point, we get $g(y_1)=G$.
Thus, keeping $x=x_1$, in problem (\ref{genpr2}), we can only obtain $g(y)\leq G$, by minimizing
the cost function over $y$.

Now take any different value $x_0\neq x_1$, satisfying
$f(x_0)=F$.
If the pair $(x_0, y_1)$ satisfies the first constraint
in problem (\ref{genpr1}), then it is a solution of problem (\ref{genpr1}), and hence the constraints should
hold with equality. This implies that $(x_0, y_1)$ also satisfies
the first constraint of problem (\ref{genpr2}).
If $(x_0, y_1)$ does not satisfy the first constraint in problem (\ref{genpr1}),
then  it certainly satisfies the first constraint of problem (\ref{genpr2}).
Either way, the pair $(x_0, y_1)$ makes a feasible solution of problem (\ref{genpr2})
(although possibly not optimal) and, by evaluating the cost function at this point, we get $g(y_1)=G$.

Thus, for any value $x_0$ we pick, we should always obtain an optimal solution
of problem (\ref{genpr2}) for which $g(y)\leq G$.
But any such optimal solution has to satisfy the first constraint with equality, hence
is also a solution of problem (\ref{genpr1}). In turn, this implies $g(y)= G$.
But then the original pair $(x_1,y_1)$ is a solution of problem (\ref{genpr2}),
since it is feasible and yields the same cost/reward function.

Take problem (\ref{genpr3}), and denote by $(x_3,y_3)$ one of its optimal solutions.
By Proposition \ref{circ_pr_prop1} we have $f(x_3)=F$ and $g(y_3)=G$.
Then either $h(x_3,y_3)\leq H$, which implies that $(x_3,y_3)$ is an optimal solution of 
problem (\ref{genpr2}), or $h(x_3,y_3)\geq H$ and then $(x_3,y_3)$ is an optimal solution of
problem (\ref{genpr1}). Either way, the inequality should hold with equality, and hence $(x_3,y_3)$
is an optimal solution of both problem (\ref{genpr1}) and problem (\ref{genpr2}), with
$h(x_3,y_3)=H$. But this also implies that $(x_1,y_1)$ is an optimal solution of problem (\ref{genpr3}).

(II) A similar argument can be made if we consider an optimal solution $(x_2,y_2)$ of problem (\ref{genpr2}),
such that $g(y_2)=G$.

(III) Consider an optimal solution $(x_3,y_3)$ of problem (\ref{genpr3}),
such that $h(x_3,y_3)=H$, and suppose there exists an optimal solution $(x_2,y_2)$ of problem
(\ref{genpr2}) is such that $g(y_2)\neq G$.
By Proposition \ref{circ_pr_prop1}, $(x_2,y_2)$ satisfies $f(x_2)=F$ and $h(x_2,y_2)=H$.
If $g(y_2)<G$, then $(x_2,y_2)$ is an optimal solution of problem (\ref{genpr3}) which
does not satisfy the constraints with equality, and thus Proposition \ref{circ_pr_prop1}
is contradicted.
If $g(y_2)=G_2>G$, then if we construct a modified version of problem (\ref{genpr3}),
where the constraint $g(y)\leq G$ is replaced by $g(y)\leq G_2$, we know by the first part of
this proof that $(x_2,y_2)$ is an optimal solution of this new problem, yielding
$h(x_2,y_2)=H$. But the same objective is attained by $(x_3,y_3)$, and moreover
$(x_3,y_3)$ satisfies the new problem's constraints since $g(y_3)=G<G_3$, and thus
is an optimal solution.
However, one of the constraints is satisfied with strict inequality, thus contradicting
Proposition \ref{circ_pr_prop1}.
Therefore, $(x_3,y_3)$ has to be a solution of problem (\ref{genpr2}).
A similar argument can be made to prove it is also a solution of problem (\ref{genpr1}).
\end{proof}


\section{Average Power Constraints: Mixed Strategies - A special two-player, zero-sum game with mixed strategies.}\label{app8}

In this section, we present a general form of a special two-player, zero-sum game with mixed strategies.
Particular forms of this game have been investigated by other authors over the last three decades.
The first simplified version was presented by Bell and Cover \cite{bell}, and a slightly
more general form was later solved by Hughes and Narayan \cite{hughes}.

\vspace*{4pt}
{\bf \emph{Problem Statement}}
\vspace*{4pt}

Let $g(y):\mathbb{R}_+ \to \mathbb{R}_+$ be a monotone increasing, almost everywhere (a.e.) continuous function
such that $g(0)=0$.
For any point of discontinuity $y_0$ such that $g(y_0^-)=x_1$ and $g(y_0^+)=x_2>x_1$, we define $g(y_0)=x_1$
($g$ is left-continuous) and $g^{-1}(x)=y_1$ for all $x\in [x_1,x_2]$.
For any interval of non-zero measure $(y_1, y_2)$ where $g$ is constant, i.e. $g(y)=x_0$ for all $y\in (y_1,y_2)$,
we define $g^{-1}(x_0)=y_1$ ($g^{-1}$ is also left-continuous). On the rest of $\mathbb{R}_+$, where $g$ is continuous
and strictly increasing, $g^{-1}$ is defined as the usual inverse function of $g$.
Note that $g^{-1}$ is a monotone increasing, a.e. continuous function.

Consider the two-player, zero-sum game with mixed strategies defined as follows. The allowable strategies for
Player 1 are all non-negative, real-valued random variables $X$ satisfying $\expec[X]\leq a$.
The allowable strategies for Player 2 are all non-negative, real-valued
random variables $Y$ satisfying $\expec[Y]\leq b$. The payoff function is
$Pr\{X\geq g(Y)\}$, which Player 1 seeks to maximize, while Player 2 seeks to minimize,
by properly picking the probability distributions of $X$ and $Y$ respectively.
Throughout the sequel, these probability distributions will be represented by their corresponding
cumulative distribution functions (CDFs) $F_X^0(x)$ and $F_Y^0(y)$.

\vspace*{4pt}
{\bf \emph{Problem Solution}}
\vspace*{4pt}

\begin{thm}\label{thm_gp1_1}
(I) If there exists a solution with $k_x,k_y\in [0,1]$ and $v\in [\max \{b/2, g^{-1}(a)/2 \}, \infty)$ of the following three equations:
\be\label{gp1_03}
k_x\left(1-\frac{b}{2v} \right)=1-k_y\left(1-\frac{a}{g(2v)} \right),
\ee
\be\label{gp1_04}
k_x=\frac{2va}{\int_{0}^{2v}g(y)dy},
\ee
\be\label{gp1_05}
k_y=\frac{g(2v)b}{\int_{0}^{g(2v)}g^{-1}(x)dx}.
\ee
then this solution is unique and the unique Nash equilibrium of the two-player, zero-sum game
described above is attained by the pair of strategies
$\left(F_X^0(x),F_Y^0(y)\right)$ satisfying:
\be\label{gp1_01}
F_X^0(g(y))\sim k_x\mathbb{U}([0,2v])(y)+(1-k_x)\Delta_0(y),
\ee 
\be\label{gp1_02}
F_Y^0(g^{-1}(x))\sim k_y\mathbb{U}([0,g(2v)])(x)+(1-k_y)\Delta_0(x),
\ee 
where $\mathbb{U}([r,t])(\cdot)$ denotes the CDF of a uniform distribution
over the interval $[r,t]$, and $\Delta_0(\cdot)$ denotes the CDF of a Dirac distribution (i.e. a step function).
\vspace*{4pt}

(II) If $g$ is strictly increasing and
continuous on $[\max \{b/2, g^{-1}(a)/2 \}, \infty)$, and $\int_0^{b}g(y)dy<\int_{g(b)}^{\infty}g^{-1}(x)dx$, then
the system in (\ref{gp1_03}), (\ref{gp1_04}) and (\ref{gp1_05}) has a unique solution such that
$k_x, k_y \in [0,1]$ and $v\in [\max \{b/2, g^{-1}(a)/2 \}, \infty)$.
Moreover, the parameters $k_p,k_j$ and $v$ are uniquely
determined from the following steps:
\begin{enumerate}
\item Find the unique value $v_0$ which satisfies:
\be
ab=[g(2v_0)-a](2v_0-b).
\ee
\item Compute $S(v_0)=\int_{0}^{2v_0}g(y)dy-2v_0 a$.
\item If $S(v_0)<0$, then $v$ is the unique solution of 
\be
\int_{0}^{2v}g(y)dy-2va=0,
\ee
\be
k_p=1
\ee
and 
\be
k_j=\frac{bg(2v)}{2v[g(2v)-a]}.
\ee
\item If $S(v_0)=0$ then $v=v_0$, $k_p=k_j=1$.
\item If $S(v_0)>0$, then $v$ is the unique solution of 
\be
\int_{0}^{2v}g(y)dy-g(2v)(2v-b)=0,
\ee
\be
k_p=\frac{2va}{g(2v)(2v-b)}
\ee
and 
\be
k_j=1.
\ee
\end{enumerate}.
\end{thm}
\vspace*{4pt}

\begin{proof}

Before starting the actual proof, several remarks are in order.
First, $F_X^0(x)$ can be computed from $F_X^0(g(y))$ by writing $x=g(g^{-1}(x))$, and thus by evaluating
$F_X^0(g(y))$ in $y=g^{-1}(x)$. A similar algorithm works for computing $F_Y^0(y)$ from $F_Y^0(g^{-1}(x))$. 

Second, note that by following this algorithm, for any point of discontinuity $y_0$ of $g$
such that $g(y_0^-)=x_1$ and $g(y_0^+)=x_2>x_1$, we have:
\be
F_X^0(x_1)=F_X^0(g(g^{-1}(x_1)))=F_X^0(g(y_0))={}\nonumber\\
{}=F_X^0(g(g^{-1}(x_2)))=F_X^0(x_2),
\ee
i.e. Player 1 does not allow $X$ to take values in $(x_1,x_2)$, and
\be
F_Y^0(y_0)=F_Y^0(y_0^+)=F_Y^0(g^{-1}(g(y_0^+)))={}\nonumber\\
{}=F_Y^0(g^{-1}(x_2)),
\ee
while by the same rational $F_Y^0(y_0^-)=F_Y^0(g^{-1}(x_1))$, meaning that Player 2 uses
a probability mass point in $y_0$.

Third, for an interval of non-zero measure $(y_1, y_2)$ where $g$ is constant, i.e. $g(y)=x_0$ for all $y\in (y_1,y_2)$,
we have:
\be
F_Y^0(y_1)=F_Y^0(g^{-1}(g(y_1)))=F_Y^0(g^{-1}(x_0))={}\nonumber\\
{}=F_Y^0(g^{-1}(g(y_2)))=F_Y^0(y_2),
\ee
i.e. Player 2 does not allow $Y$ to take values in $(y_1,y_2)$, and
\be
F_X^0(x_0)=F_X^0(x_0^+)=F_X^0(g(g^{-1}(x_0^+)))={}\nonumber\\
{}=F_X^0(g(y_2)),
\ee
while by the same rational $F_X^0(x_0^-)=F_X^0(g(y_1))$, meaning that Player 1 uses
a probability mass point in $x_0$.
We now proceed with the proof of the first part of the theorem.

(I) Since this is a two-player, zero-sum game with mixed strategies, it has a unique Nash equilibrium.
Let $X_0\sim F_X^0$ and $Y_0\sim F_Y^0$ denote the random variables with the CDFs in (\ref{gp1_01}) and (\ref{gp1_02}), and
$X\sim F_X$ and $Y\sim F_Y$ be any arbitrary random variables.

Note that $Pr\{X\geq g(Y)\}=\int_{0}^{\infty}[1-F_X(g(y))]dF_Y(y)=\int_{0}^{\infty}F_Y(g^{-1}(x))dF_X(x)$.
We can write
\be\label{gp1_1}
Pr\{X_0\geq g(Y)\}=\int_{0}^{\infty}[1-F_X^0(g(y))]dF_Y(y)={}\nonumber\\
{}=1-k_x\int_{0}^{\infty}\mathbb{U}([0,2v])(y)dF_Y(y)-{}\nonumber\\
{}-(1-k_x)\int_{0}^{\infty}\Delta_0(y)dF_Y(y)\geq{}\nonumber\\
{}\geq k_x\left(1-\frac{1}{2v}\int_{0}^{\infty}ydF_Y(y) \right) \geq k_x\left(1-\frac{b}{2v}\right),
\ee
and
\be\label{gp1_2}
Pr\{X\geq g(Y_0)\}=\int_{0}^{\infty}F_Y^0(g^{-1}(x))dF_X(x)={}\nonumber\\
{}=k_y\int_{0}^{\infty}\mathbb{U}([0,g(2v)])(x)dF_X(x)+{}\nonumber\\
{}+(1-k_y)\int_{0}^{\infty}\Delta_0(x)dF_X(x)\leq{}\nonumber\\
{}\leq 1-k_y\left(1-\frac{1}{g(2v)}\int_{0}^{\infty}xdF_X(x) \right) \leq{}\nonumber\\
{}\leq 1-k_y\left(1-\frac{a}{g(2v)}\right).
\ee

Note that equality holds in the first inequality of (\ref{gp1_1}) if $F_Y(2v)=1$, and in the second
inequality of (\ref{gp1_1}) if $\expec[Y]=b$. Similarly, equality holds in the first inequality of
(\ref{gp1_2}) if $F_X(g(2v))=1$, and in the second inequality of (\ref{gp1_2}) if $\expec[X]=a$.

Since $F_Y^0(2v)=F_Y^0(g^{-1}(g(2v)))=1$ and $F_X^0(g(2v))=1$ (see (\ref{gp1_01}), (\ref{gp1_02})),
equalities hold in (\ref{gp1_1}) and (\ref{gp1_2}) when $F_X=F_X^0$ and $F_Y=F_Y^0$ if and only if
\be\label{rela1}
a=\int_{0}^{\infty}xdF_X^0(x)
\ee
and
\be\label{relb1}
b=\int_{0}^{\infty}ydF_Y^0(y).
\ee 
Although the two CDFs $F_X^0(x)$ and $F_Y^0(y)$ may not be continuous as functions in $\mathscr{L}_1$, they admit derivatives
in the distribution space $\mathscr{D}'$ \cite{zeman}, and thus we can write
\be
\int_{0}^{\infty}xdF_X^0(x)=\int_{0}^{\infty}x\frac{dF_X^0(x)}{dx}dx={}\nonumber\\
{}=\int_{0}^{\infty}g(y)\frac{dF_X^0(g(y))}{dg(y)} \frac{dg(y)}{dy}dy={}\nonumber\\
{}=\int_{0}^{\infty}g(y)\frac{dF_X^0(g(y))}{dy}dy={}\nonumber\\
{}=(1-k_x)\int_{0}^{\infty}\delta_0(y)g(y)dy+\frac{k_x}{2v}\int_{0}^{\infty}g(y)dy,
\ee
which along with (\ref{rela1}) results in (\ref{gp1_04}), and similarly
\be
\int_{0}^{\infty}ydF_Y^0(y)
=\int_{0}^{\infty}g^{-1}(x)\frac{dF_Y^0(g^{-1}(x))}{dx}dx={}\nonumber\\
{}=(1-k_y)\int_{0}^{\infty}\delta_0(x)g^{-1}(x)dx+{}\nonumber\\
{}+\frac{k_y}{g(2v)}\int_{0}^{\infty}g^{-1}(x)dx,
\ee
which together (\ref{relb1}) yields (\ref{gp1_05}).
The conditions for $\left(F_X^0(x),F_Y^0(y)\right)$ to achieve a saddle-point is that equality
holds between the bounds in (\ref{gp1_1}) and (\ref{gp1_2}), which translates to (\ref{gp1_03}), and that
there always exists a solution of the system given by (\ref{gp1_03}), (\ref{gp1_04}) and (\ref{gp1_05}).

(II) This part of the theorem provides a general (although not necessary) condition for such a solution to
exist and states that under this condition no more than one such a solution can exist (although the
uniqueness already follows as a consequence of the uniqueness of a Nash equilibrium).
By substituting (\ref{gp1_04}) and (\ref{gp1_05}) in (\ref{gp1_03}) we get
\be\label{gp1_3}
\frac{a(2v-b)}{\int_0^{2v}g(y)dy}=1-\frac{b(g(2v)-a)}{\int_0^{g(2v)}g^{-1}(x)dx}.
\ee
Denote the left hand side of (\ref{gp1_3}) by $L(v)$ and the right hand side by $R(v)$
for simplicity.
Note that for any function $g$ that satisfies the conditions set in the problem formulation we have
\be\label{gp1_4}
\int_0^{2v}g(y)dy=2vg(2v)-\int_0^{g(2v)}g^{-1}(x)dx.
\ee
This relation is best observed graphically in Figure \ref{gyfig2}.
\begin{figure}[h]
\centering
\includegraphics[scale=1.0]{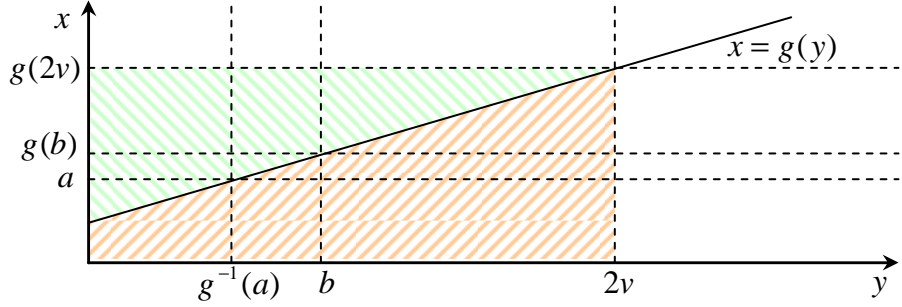}
\caption{The relationship between the integrals of $g(y)$ and $g^{-1}(x)$.}\label{gyfig2}
\end{figure}

Computing the derivatives of $L(v)$ and $R(v)$ with respect to $v$
(these derivatives always exist for $v\geq \max \{b/2, g^{-1}(a)/2 \}$) we get
\be\label{gp1_5}
\frac{dL(v)}{dv}=\frac{2a}{\left[\int_0^{2v}g(y)dy\right]^2}\cdot{}\nonumber\\
{}\cdot\left[\int_0^{2v}g(y)dy-g(2v)(2v-b)\right],
\ee
and
\be\label{gp1_6}
\frac{dR(v)}{dv}=\frac{2g'(v)b}{\left[\int_0^{g(2v)}g^{-1}(x)dx\right]^2}\cdot{}\nonumber\\
{}\cdot\left[2v(g(2v)-a)-\int_0^{g(2v)}g^{-1}(x)dx\right]={}\nonumber\\
{}=\frac{2g'(v)b}{\left[\int_0^{g(2v)}g^{-1}(x)dx\right]^2}\left[\int_0^{2v}g(y)dy-2va\right],
\ee
where $g'(v)>0$ denotes the first derivative $dg(y)/dy$, evaluated in $y=v$, and 
the second equality in (\ref{gp1_6}) follows from (\ref{gp1_4}).

Note that $L(v)$ and $R(v)$ are both probabilities, hence belong to $[0,1]$.
Therefore, any possible solution of the system in (\ref{gp1_03}), (\ref{gp1_04}) and (\ref{gp1_05})
should satisfy $2v\geq b$ and $g(2v)\geq a$, or equivalently:
\be\label{gp1_6_1}
v\geq \max \{b/2, g^{-1}(a)/2 \}.
\ee
Therefore, in the sequel of this proof we shall implicitly assume that (\ref{gp1_6_1})
holds true.

Denote $S_L(v)=\int_0^{2v}g(y)dy-g(2v)(2v-b)$ and $S_R(v)=\int_0^{2v}g(y)dy-2va$.
Since
\be
\frac{d}{dv}\int_0^{2v}g(y)dy=2g(2v),
\ee
we observe that
\be
\frac{d}{dv}S_L(v)=-2g'(v)(2v-b)<0
\ee
and
\be
\frac{d}{dv}S_R(v)=2(g(2v)-a)>0,
\ee
which imply that $S_L(v)$ is a strictly decreasing function of $v$, while $S_R(v)$ is a strictly
increasing function of $v$, for
the domain of interest $v\in [\max \{b/2, g^{-1}(a)/2 \}, \infty)$.

Note that $\frac{d}{dv}S_R(v)$ is strictly positive even in the limit as $v\to\infty$, and thus
$\lim_{v\to\infty}S_R(v)=\infty$.
By writing $S_L(v)=\int_0^{b}g(y)dy-\int_{g(b)}^{g(2v)}g^{-1}(x)dx$,
we also have $\lim_{v\to\infty}S_L(v)=-\infty$.

\emph{A first possible solution:}

An extremum of $L(v)$ is obtained by setting $\frac{dL(v)}{dv}=0$, or equivalently
\be\label{gp1_7}
\int_0^{2v_l}g(y)dy=g(2v_l)(2v_l-b).
\ee
In our previously introduced notation, this writes $S_L(v_l)=0$.
But since $S_L(v)$ is strictly decreasing on the domain of interest, the extremum is unique and is a maximum.

The values of $L(v)$ and $R(v)$ at this point are given by
\be\label{gp1_sol1}
L(v_l)=R(v_l)=\frac{a}{g(2v_l)}.
\ee
Moreover, substituting (\ref{gp1_7}) and (\ref{gp1_4}) back in (\ref{gp1_04}) and (\ref{gp1_05}) we get
\be\label{gp1_8}
k_{x,l}=\frac{2v_la}{g(2v_l)(2v_l-b)}
\ee
and
\be\label{gp1_9}
k_{y,l}=1.
\ee
Therefore $(v_l, k_{x,l}, k_{y,l})$ are a solution of the system given by
(\ref{gp1_03}), (\ref{gp1_04}) and (\ref{gp1_05}) if and only if $k_{x,l}\in [0,1]$.
From (\ref{gp1_6_1}) it is implied that $2v_l\geq b$, and hence that $k_{x,l}\geq0$.
The condition $k_{x,l}\leq 1$ yields
\be\label{gp1_10}
2v_la\leq g(2v_l)(2v_l-b).
\ee

\emph{A second possible solution:}

An extremum of $R(v)$ is obtained by setting $\frac{dR(v)}{dv}=0$, or equivalently
\be\label{gp1_11}
\int_0^{2v_r}g(y)dy=2v_ra.
\ee
In our previously introduced notation, this writes $S_R(v_r)=0$.
When this extremum of $R(v)$ exists, it is also unique and is a minimum,
since $S_R(v)$ is strictly increasing on the domain of interest. 

The values of $L(v)$ and $R(v)$ at this point are given by
\be\label{gp1_sol2}
L(v_r)=R(v_r)=1-\frac{b}{2v_r}.
\ee
Moreover, substituting (\ref{gp1_11}) back in (\ref{gp1_04}) and (\ref{gp1_05}) we get
\be\label{gp1_12}
k_{x,r}=1
\ee
and
\be\label{gp1_13}
k_{y,r}=\frac{bg(2v_r)}{2v_r(g(2v_r)-a)}.
\ee
Therefore $(v_r, k_{x,r}, k_{y,r})$ are a solution of the system given by
(\ref{gp1_03}), (\ref{gp1_04}) and (\ref{gp1_05}) if and only if $k_{y,r}\in [0,1]$.
From (\ref{gp1_6_1}) it is implied that $g(2v_r)\geq a$, and hence that $k_{y,r}\geq0$.
The condition $k_{y,r}\leq 1$ yields the same inequality as before:
\be
2v_ra\leq g(2v_r)(2v_r-b).
\ee

Recall that $L(v)$ has a unique maximum, while $R(v)$ has a unique minimum.
The immediate implication of this is that the equation $L(v)=R(v)$ can have a maximum of two solutions.
These are the possible solutions discussed above.

To summarize, have two sets of relations:
\be\label{gp1_14}
\int_0^{2v_l}g(y)dy=g(2v_l)(2v_l-b),\nonumber\\
2v_la\leq g(2v_l)(2v_l-b)
\ee
and
\be\label{gp1_15}
\int_0^{2v_r}g(y)dy=2v_ra,\nonumber\\
2v_ra\leq g(2v_r)(2v_r-b)
\ee 
that could each yield a solution of the system in (\ref{gp1_03}), (\ref{gp1_04}) and (\ref{gp1_05}).

In the remainder of this proof, we show that at least one of the sets (\ref{gp1_14})
and (\ref{gp1_15}) has a solution and the sets
(\ref{gp1_14}) and (\ref{gp1_15}) cannot both have different solutions.
 
Let $v_0$ denote the value of $v$ in $[\max\{b/2, g^{-1}(a)/2\},\infty)$ for which
\be\label{gp1_16}
2v_0a=g(2v_0)(2v_0-b), as in Figure \ref{gyfig1}.
\ee
\begin{figure}[h]
\centering
\includegraphics[scale=1.0]{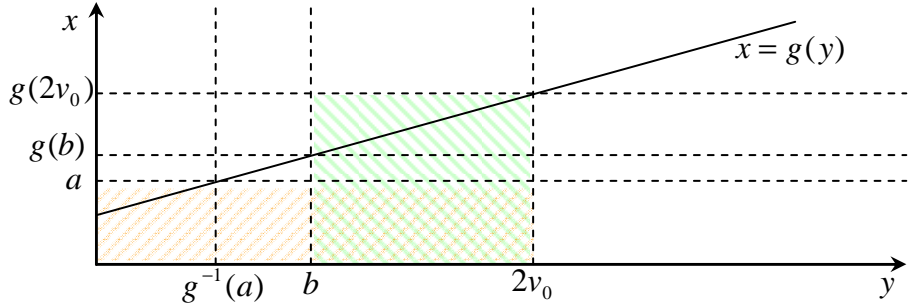}
\caption{Finding $v_0$.}\label{gyfig1}
\end{figure}

Such a value exists and is unique since (\ref{gp1_16}) is equivalent to
$ab=(g(2v_0)-a)(2v_0-b)$, where the term on the right hand side is a strictly increasing
function of $v_0$ on $[\max\{b/2, g^{-1}(a)/2\},\infty)$, with a minimum in $v_0=\max\{b/2, g^{-1}(a)/2\}$ which is $0$
and $\lim _{v\to \infty}(g(2v)-a)(2v-b)=\infty $.
Note that this also implies that $2va\leq g(2v)(2v-b)$ can only be satisfied if $v>v_0$.

Denote $S=S_L(v_0)=S_R(v_0)$ the common value of $S_L$ and $S_R$ in $v_0$.
If $S=0$, then $v_l=v_r=v_0$.
If $S<0$ or $S>0$, since $S_L(v)$ is decreasing with $v$ and $S_R(v)$ is increasing with $v$ for the domain of interest,
it is not possible to obtain solutions larger than $v_0$ to both equations $S_L(v)=0$ and $S_R(v)=0$.

However, a solution always exists. If $S<0$, the solution is guaranteed by the continuity of $S_R(v)$
on the domain of interest, and by the fact that $\lim_{v\to\infty}S_R(v)=\infty$.
If $S>0$, the solution is guaranteed by the continuity of $S_L(v)$
on the domain of interest, and by the fact that $\lim_{v\to\infty}S_L(v)<0$, which follows
from the condition $\int_0^{b}g(y)dy<\int_{g(b)}^{\infty}g^{-1}(x)dx$. Note that this condition
is only necessary if $S>0$ and is illustrated in Figure \ref{gyfig3}.

\begin{figure}[h]
\centering
\includegraphics[scale=1.0]{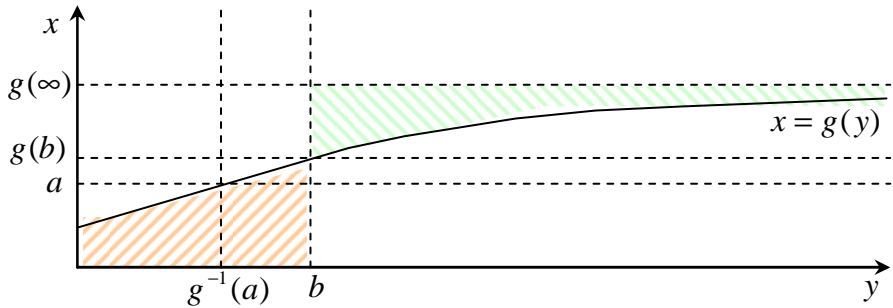}
\caption{The necessary condition for the existence of a solution when $S>0$.}\label{gyfig3}
\end{figure}

A similar condition can be written for the case when $S<0$, that is
$\lim_{v\to\infty}S_R(v)>0$ if and only if $\int_0^{a}g^{-1}(x)dx<\int_{g^{-1}(a)}^{\infty}g(y)dy$.
However, since $g$ is a function and is defined over $\mathbb{R}_+$, this latter condition can only be violated if
$g$ is constant on $[a,\infty)$. But this is impossible under the former condition.

We have thus shown that under the condition that $g$ is strictly increasing and
continuous on $[\max \{b/2, g^{-1}(a)/2 \}, \infty)$, and $\int_0^{b}g(y)dy<\int_{g(b)}^{\infty}g^{-1}(x)dx$,
the system given by (\ref{gp1_03}), (\ref{gp1_04}) and (\ref{gp1_05}) always has a solution,
and that this solution is unique.
\end{proof}

\vspace*{4pt}
{\bf \emph{Several additional remarks}}
\vspace*{4pt}

Bell and Cover \cite{bell} found the solution of our game for the particular case when $a=b=1$ and $g(y)=y$.
In the context of Gaussian arbitrarily varying channels, Hughes and Narayan \cite{hughes} extended
the previous result to the case where $a$ and $b$ are any positive constants, and $g(y)=y+c$, with
$c\geq 0$. In the remainder of this section we show that our results can be easily particularized
to obtain the same results as in \cite{hughes}.

\begin{figure}[h]
\centering
\includegraphics[scale=1.0]{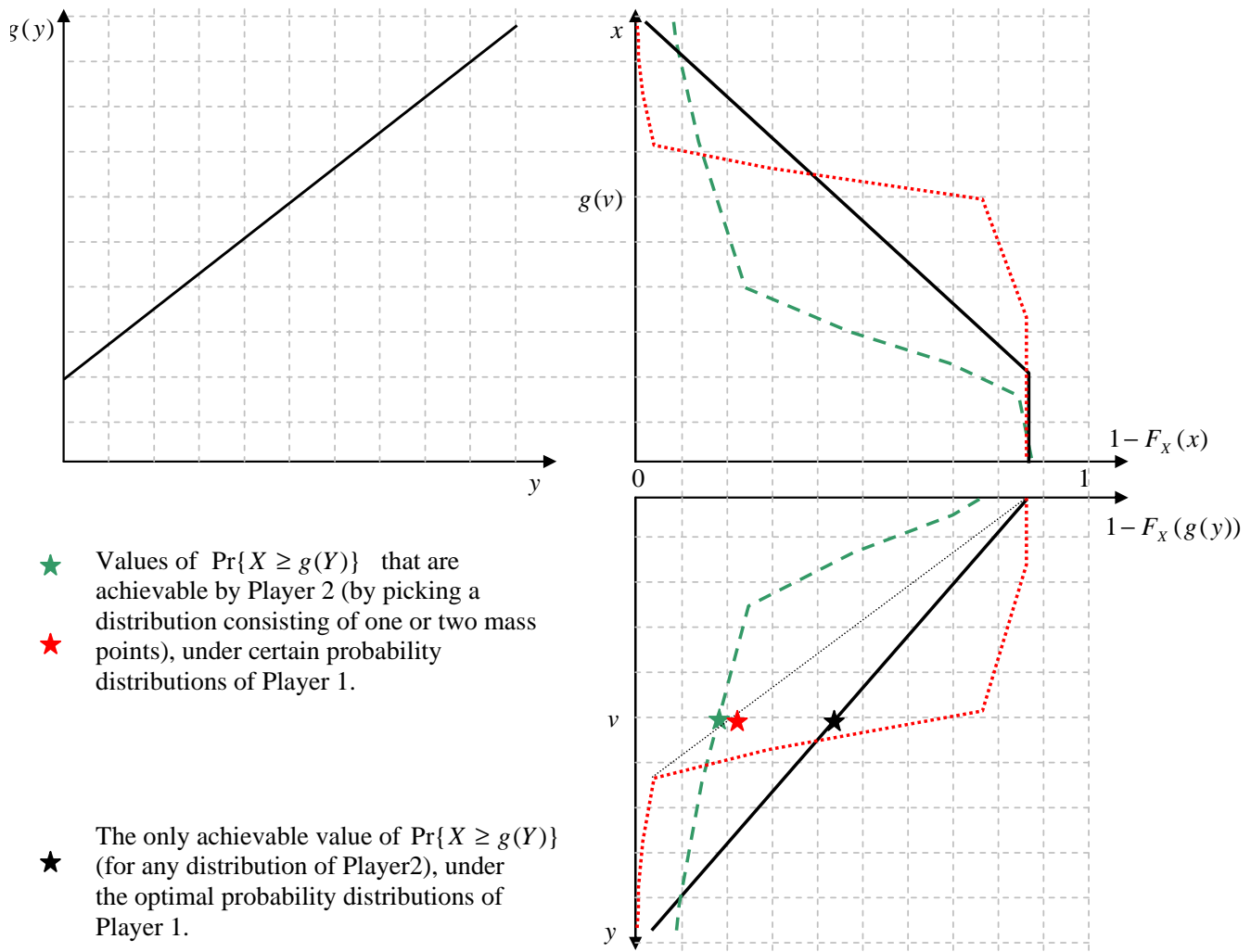}
\caption{Intuitive explanation for the optimality of the strategy in (\ref{gp1_01}).}\label{intuitiveapproach1}
\end{figure}

\begin{figure}[h]
\centering
\includegraphics[scale=1.0]{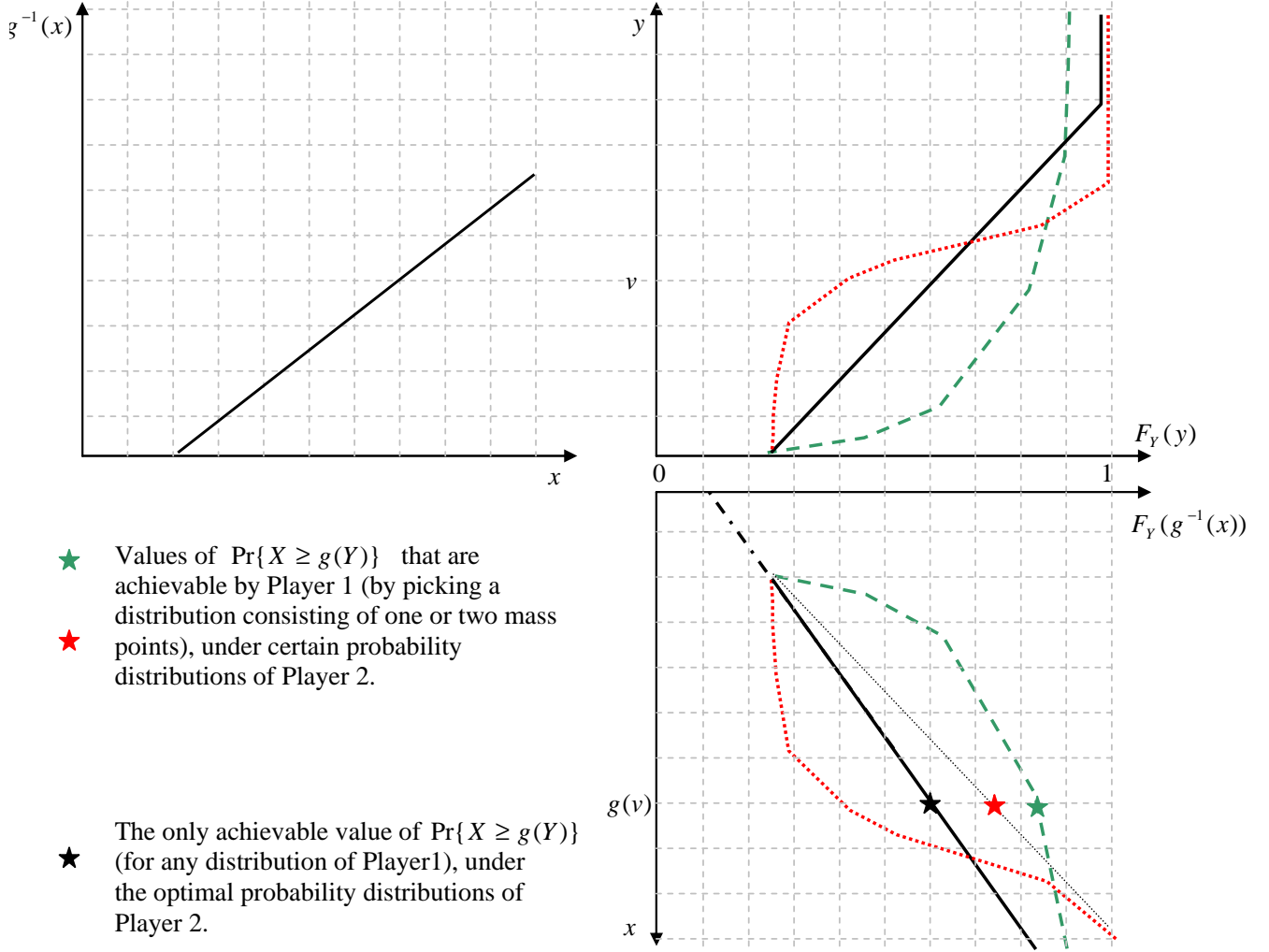}
\caption{Intuitive explanation for the optimality of the strategy in (\ref{gp1_02}).}\label{intuitiveapproach2}
\end{figure}

If we force $g(0)=g(0^-)=0$, the function $g(y)=y+c,~\forall y>0 $ is unbounded, linear, strictly increasing,
and has only one discontinuity in $y=0$. Hence, it satisfies all the conditions set in the problem formulation,
as well as those of part (II) of our Theorem \ref{thm_gp1_1}.

Substituting $g(y)=y+c$ in (\ref{gp1_14}), we get (Case 1):
\be
2v_l^2-2v_lb-bc=0
\ee
and
\be
a\leq v_l+c,
\ee
resulting in
\be
v_l=\frac{b}{2}\left[1+\sqrt{1+\frac{2c}{b}}\right]
\ee
under the condition that
\be \label{gp1_cond01}
a\leq c+\frac{b}{2}\left[1+\sqrt{1+\frac{2c}{b}}\right].
\ee

The cost function for this case results from (\ref{gp1_sol1}) as
\be
Pr\{X\geq g(Y)\}=\frac{a}{c+b\left[1+\sqrt{1+\frac{2c}{b}}\right]}={}\nonumber\\
{}=\frac{a}{c}\left[1+\frac{b}{c}\left(1-\sqrt{1+\frac{2c}{b}}\right)\right],
\ee
and is also consistent with \cite{hughes}.
Note that although $k_y=1$ for this case, this does not mean that Player 2 is always on.
Recall that a discontinuity of $g$ is translated into a mass point for the probability
distribution of $Y$. In this case, the discontinuity in $y=0$ means that $Y=0$ with
probability $\frac{c}{g(2v_l)}=1-\frac{b}{v_l}$, which is the same as in \cite{hughes}.

Similarly, substituting $g(y)=y+c$ in (\ref{gp1_15}), we get (Case 2): 
\be
v_l=\frac{b}{2}\left[1+\sqrt{1+\frac{2c}{b}}\right]
\ee
under the condition that
\be\label{gp1_cond02}
a\geq c+\frac{b}{2}\left[1+\sqrt{1+\frac{2c}{b}}\right].
\ee
Note that the two conditions (\ref{gp1_cond01}) and (\ref{gp1_cond02}) are
mutually exclusive.
The cost function for this case is
\be
Pr\{X\geq g(Y)\}=1-\frac{b}{2(a-c)},
\ee
and is consistent with \cite{hughes}.

In Figure \ref{intuitiveapproach1} we offer an intuitive explanation of why $F_X^0(g(y))$ should be uniform
over $[0,2v]$, from a maximin point of view. The function $g(y)$ is taken to be linear, with a discontinuity
in $0$, similar to \cite{hughes}.
Assuming that Player 1 plays first (maximin), we note that if $F_X^0(g(y))$ is not uniform, the second player can pick
a strategy that decreases the value of the objective $Pr\{X\geq g(Y)\}$. Therefore, in order to provide the second player
with an indifferent choice space (the strategy of Player 2 can be any probability distribution over $[0,2v]$), Player 1
should pick $F_X^0(x)$ such that $F_X^0(g(y))$ is uniform over $[0,2v]$.  

\begin{figure}[h]
\centering
\includegraphics[scale=1.0]{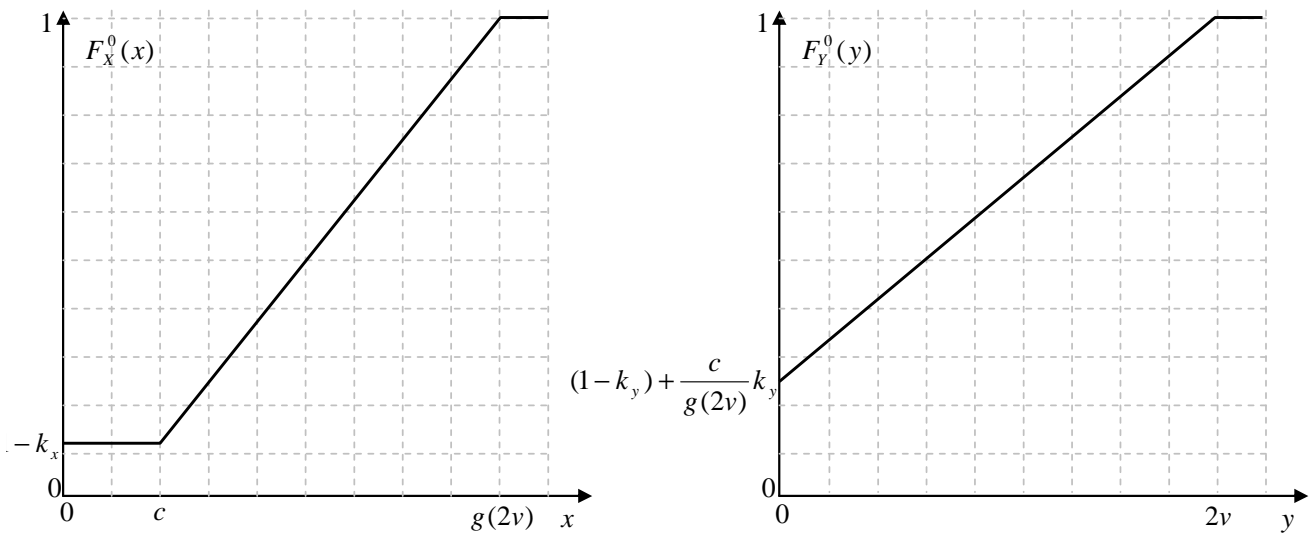}
\caption{The resulting strategies $F_X^0(x)$ and $F_Y^0(y)$ for a linear $g(y)$ with a discontinuity in $0$.}\label{intuitiveapproach3}
\end{figure}

Similarly, in Figure \ref{intuitiveapproach2} we offer an intuitive explanation of why $F_Y^0(g^{-1}(x))$ should be uniform
over $[0,g(2v)]$, from a minimax point of view.
Assuming that Player 2 plays first (minimax), note that if $F_Y^0(g^{-1}(x))$ is not uniform, the first player can pick
a strategy that increases the value of the objective $Pr\{X\geq g(Y)\}$.

The optimal distributions resulting from Figures \ref{intuitiveapproach1} and \ref{intuitiveapproach2} are shown in
Figure \ref{intuitiveapproach3}. They are consistent with our theoretical results (and the results of \cite{hughes})
for $g(y)=y+c$.

\bibliographystyle{IEEEtran}
\bibliography{jamming}
\end{document}